\documentclass[twocolumn]{svjour3}

\usepackage{svn-multi}
\svnid{$Id: amber-vldbj13.tex 4104 2012-09-25 20:22:21Z timfu $}

\usepackage{graphicx}
\usepackage[T1]{fontenc}
\makeatletter
\svnid{$Id: _macros.tex 4155 2012-09-26 10:18:57Z schallhart $}

\usepackage{graphicx}
\usepackage{paralist}
\usepackage{mathptmx}
\usepackage{amsmath}
\usepackage{color}
\usepackage{multirow}
\usepackage[T1]{fontenc}

\DeclareMathAlphabet{\mathcal}{OMS}{cmsy}{m}{n}

\newcommand{\algoname}[1]{\texttt{#1}\xspace}
\newcommand{\annotate}{\algoname{annotate}}
\newcommand{\learn}{\algoname{learn}}
\newcommand{\amberalgo}{\algoname{amber}}
\newcommand{\dataareaalgo}{\algoname{identify}}
\newcommand{\recordalgo}{\algoname{segment}}
\newcommand{\attributealgo}{\algoname{align}}
\newcommand{\learningalgo}{\algoname{learn}}

\newcommand{\partOf}{{\normalfont\textsf{partOf}}\xspace}
\newcommand{\TYPING}{\ensuremath{\mathcal{T}}\xspace}
\newcommand{\dataareatype}{\ensuremath{\mathbf{d}}\xspace}
\newcommand{\recordstarttype}{\ensuremath{\mathbf{rs}}\xspace}
\newcommand{\recordtailtype}{\ensuremath{\mathbf{rt}}\xspace}

\newcommand{\UNIVERSE}{\ensuremath{U}\xspace}
\newcommand{\DOMNODES}{\ensuremath{N}\xspace}

\newcommand{\SCHEMA}{\ensuremath{\Sigma}\xspace}
\newcommand{\OPTATTS}{\ensuremath{\Sigma_O}\xspace}
\newcommand{\REGATTS}{\ensuremath{\Sigma_R}\xspace}

\newcommand{\Partition}{\ensuremath{\mathcal{R}}\xspace}
\newcommand{\OptimalPartition}{\ensuremath{\mathcal{R}_{opt}}\xspace}

\newcommand{\RECORDS}{\ensuremath{\mathcal{R}}\xspace}

\newcommand{\ATYPE}[1]{\textsf{\textsc{#1}}\xspace}

\newcommand{\DOMREL}[1]{{\normalfont\textsf{#1}}\xspace}

\newcommand{\DEPTH}{{\normalfont\textsf{depth}}}
\newcommand{\editDistance}{{\normalfont\textsf{editDist}}}
\newcommand{\irregularity}{{\normalfont\textsf{irregularity}}}

\renewenvironment{compactenum}[1][]{\begin{enumerate}[#1]}{\end{enumerate}}

\newcommand{\deskip}[1]{}

\newcommand{\smalldeskip}{\deskip{.4}}

\newcommand{\nop}[1]{}

\newcommand{\AMBER}{\sys}
\newcommand{\AMBERnofont}{AMBER\xspace}
\newcommand{\sys}{\tool{Amber}}
\newcommand{\ROADRUNNER}{\tool{RoadRunner}}
\newcommand{\MDR}{\tool{MDR}}
\newcommand{\tool}[1]{\textsc{#1}\xspace}

\newcommand{\htmltag}[1]{{\normalfont\texttt{#1}}\xspace}
\newcommand{\intMerge}{\uplus}
\newcommand{\unary}{\ensuremath{\textsf{unary}}\xspace}
\newcommand{\fctn}[1]{\textsf{#1}}
\newcommand{\setfctn}[1]{\textsf{\textsc{#1}}}

\newcommand{\ANNOT}{\fctn{ann}\xspace}
\newcommand{\PIVOTS}{\setfctn{pivots}\xspace}
\newcommand{\LEADINGS}{\setfctn{leadings}}

\newcommand{\THRES}[1]{\ensuremath{\Theta^{\fctn{#1}}}\xspace}
\newcommand{\DEPTHTHRES}{\THRES{depth}}
\newcommand{\DISTTHRES}{\THRES{dist}}
\newcommand{\InventThres}{\THRES{infer}}
\newcommand{\DeleteThres}{\THRES{keep}}

\newcommand{\OptionalInventThres}{\THRES{infer}_{O}}
\newcommand{\OptionalDeleteThres}{\THRES{keep}_{O}}
\newcommand{\RegularInventThres}{\THRES{infer}_{R}}
\newcommand{\RegularDeleteThres}{\THRES{keep}_{R}}

\newcommand{\PRECEDES}{\ensuremath{\prec}\xspace}
\newcommand{\PRECEDESEQ}{\ensuremath{\preceq}\xspace}

\newcommand{\type}[1]{\textsf{\textsc{#1}}}
\newcommand{\TYPE}[1]{\textsf{\textsc{#1}}}
\newcommand{\SDIST}{\ensuremath{-_{\textsf{sibl}}}}
\newcommand{\TAGPATH}{\normalfont\textsf{tag-path}}
\newcommand{\SUPPORT}{\normalfont\textsf{supp}}

\newenvironment{compactdef}{\smalldeskip\begin{definition}}%
        {\end{definition}\smalldeskip}
\newenvironment{compacttheo}{\smalldeskip\begin{theorem}}%
        {\end{theorem}\smalldeskip}
        {\end{proposition}\smalldeskip}
\newenvironment{compactproof}{\smalldeskip\begin{proof}}%
        {\end{proof}\smalldeskip}

\usepackage{url}
\usepackage{xspace}

\usepackage{hhline}
\usepackage{amsmath}
\usepackage{amssymb}
\usepackage{stmaryrd}
\usepackage{booktabs}
\usepackage[nounderscore]{syntax}
\graphicspath{{figures/}{images/}}
  \DeclareGraphicsExtensions{.pdf,.jpeg,.png}
\usepackage{subfig}
\usepackage{listings}
\usepackage{xcolor}
\lstdefinelanguage{Xcerpt}{%
      morekeywords=[1]{GOAL,CONSTRUCT,FROM,END,DECLARE,PROGRAM},
      morekeywords=[2]{declare,all,some,position,pos,optional,opt,where,order-by,group-by,and,or,not},%
      morekeywords=[2]{variable,ns-default,ns-prefix,not,if,then,else,case,without,desc,descendant,in,identity},%
      morekeywords=[3]{idvar,var,->,as},
      morekeywords=[4]{lexical,sum,count,join},
      morekeywords=[5]{},
      morestring=[b]",%
      morecomment=[l]{\#},%
      morecomment=[s]{\\\#}{\#\\\\)},%
      alsodigit={-},%
      literate= {->}{$\rightarrow\;$}{2},%
      sensitive%
    }[keywords,strings]

    \lstdefinelanguage{XQuery}{%
      morekeywords=[1]{for,in,let,where,return,if,then,else,case,satisfies,default,typeswitch},%
      morekeywords=[1]{child,descendant,attribute,self,descendant-or-self,following-sibling,following,parent,ancestor,preceding-sibling,preceding,ancestor-or-self,},%
      morekeywords=[2]{cast,as,castable,instance,of,some,every,in,unordered,ordered,stable,order,by,descending,empty,collation,derives-from,nilled,nillable,node,schema-element,document-node},%
      morekeywords=[3]{element,function,variable,comment,processing-instruction,text,document,attribute,to},%
      morekeywords=[4]{xquery,declare,namespace,module,option,ordering,version,encoding,order,last,copy-namespaces,preserve,no-inherit,import,schema,at,external,construction,strip,base-uri,boundary-space},%
      morekeywords=[3]{fn:count,fn:sum,fn:empty,fn:max,fn:doc,fn:avg,fn:node-name,fn:last,fn:position,fn:error,position,last,size,count,substring,fn:substring,string,fn:string,contains,fn:contains,concat,fn:concat,substring-after,fn:substring-after,fn:string-join,string-join},%
      morekeywords=[5]{mod,eq,or,and,idiv,true,false,is,div,union,intersect,except,gt,lt,},%
      morestring=[b]",%
      morecomment=[s]{(:}{:)},%
      alsoletter={-},
      sensitive%
    }[keywords,comments,strings]

    \lstdefinelanguage{XPath1}{%
      morekeywords=[1]{child,descendant,attribute,self,descendant-or-self,following-sibling,following,parent,ancestor,preceding-sibling,preceding,ancestor-or-self,desc,foll,prec},%
      morekeywords=[2]{comment,processing-instruction,text,attribute},%
      morekeywords=[3]{fn:count,fn:sum,fn:empty,fn:max,fn:doc,fn:avg,fn:node-name,fn:last,fn:position,fn:error,position,last,size,count,substring,fn:substring,string,fn:string,contains,fn:contains,concat,fn:concat,boolean},%
      morekeywords=[5]{or,and,true,false,union,intersect,except},%
      morestring=[b]",%
      morestring=[b]',%
      morecomment=[s]{(:}{:)},%
     alsoletter={-},
      literate={<----}{$\longleftarrow\;\,\,$}{3},
      sensitive%
    }[keywords,comments,strings]

    \lstdefinelanguage{OXPath}[]{XPath1}{%
      morekeywords=[1]{style},%
      morekeywords=[2]{field,any-field},%
      morekeywords=[3]{},%
      morekeywords=[4]{},%
      morekeywords=[5]{},%
      morecomment=[s]{\{}{\}},%
	commentstyle=\bfseries\color{red},
   }[keywords,comments,strings]

    \lstdefinelanguage{CSS}{%
      morekeywords=[1]{>,+,*,\#,|,~,not,nth-child,attribute,first-of-type,first-line,last-of-type,last-child,empty,nth-of-type},%
      morekeywords=[2]{@import,@media,@charset},
morekeywords=[3]{color,font-weight,margin,margin-left,padding-left,border-left,background-color,font-family,text-transform,font-size,border,text-align,background,font-style},%
      morekeywords=[4]{sans-serif,rgb,solid,none,normal,uppercase,bold,print},%
      morestring=[b]",%
     alsoletter={>,+,\#,|,~,-,@},
     alsoother={:,::},
     sensitive%
    }[keywords,comments,strings]

    \lstdefinelanguage{XMLSchema}[]{XML}{
      morekeywords=[1]{xs:complexType,xs:element,xs:sequence,xs:choice,xs:simpleType,xs:key,xs:selector,xs:unique,xs:keyref,xs:field,xs:restriction,xs:minExclusive,xs:attribute,xs:extension,xs:complexContent,xs:schema},
      morekeywords=[3]{base,name,minOccurs,maxOccurs,type,xpath,refer,use}
    }
    \lstdefinelanguage{FancyXML}[]{XML}{
      morecomment=[s]{<}{>},%
	commentstyle=\bfseries\color{darkblue},
}
    \lstdefinelanguage{RNC}{%
      morekeywords=[1]{start,element,attribute,text,empty},
      morekeywords=[2]{default,namespace},%
      morekeywords=[4]{grammar,include,parent},%
      morekeywords=[3]{},
      morekeywords=[5]{},
      morestring=[b]",%
      morecomment=[l]{\#},%
      alsodigit={-},%
      sensitive%
    }[keywords,comments,strings]

\lstdefinelanguage{SPARQL}{%
  morekeywords=[1]{CONSTRUCT,WHERE,SELECT},
  morekeywords=[2]{AND,FILTER,UNION,OPT,OPTIONAL,MINUS},%
  morekeywords=[3]{sameTerm,isBLANK,isLITERAL,isIRI,BOUND},
  morekeywords=[4]{},
  morekeywords=[5]{},
  morestring=[b]",%
  alsodigit={-},%
}[keywords,strings]

\lstdefinelanguage{TurtleTwo}[]{SPARQL}{%
  morekeywords=[1]{@prefix},
  morekeywords=[2]{rdf,rdfs,fb},%
  morekeywords=[5]{type,subClassOf,domain,range,subPropertyOf},
  morekeywords=[6]{hasPart,linksTo,title,genre,author},
}[keywords,strings]

    \definecolor{lightblue}{rgb}{0,0,.7}
    \definecolor{orange}{rgb}{1,.7,0}
    \definecolor{darkorange}{rgb}{1,.4,0}
    \definecolor{darkgreen}{rgb}{0,.8,0}
    \definecolor{darkblue}{rgb}{0,0,.4}
    \definecolor{darkred}{rgb}{.4,0,0}
    \definecolor{gray}{rgb}{.2,.2,.2}
    \definecolor{darkgray}{rgb}{.4,.4,.4}
    \definecolor{shadecolor}{gray}{0.975}

\definecolor{purple}{rgb}{0.65, 0.12, 0.82}
\definecolor{flexred}{rgb}{0.65, 0.01, 0.01}
\definecolor{flexgreen}{rgb}{0, 0.6, 0}
\definecolor{flexgray}{rgb}{0.25, 0.37, 0.75}
\definecolor{flexblue}{rgb}{0, 0.2, 1}
\definecolor{flexfunction}{rgb}{0.2, 0.6, 0.4}
\definecolor{flexvar}{rgb}{0.4, 0.6, 0.8}

\lstdefinelanguage{JavaScript} {
	sensitive=true,
	morecomment=[l][\color{flexgreen}\ttfamily]{//},
	morecomment=[s][\color{flexgreen}\ttfamily]{/*}{*/},
	morecomment=[s][\color{flexgray}\ttfamily]{/**}{*/},
	morestring=[b]",
	stringstyle=\color{flexred}\textbf,
	commentstyle=\color{flexgreen},
	showstringspaces=false,
	numberstyle=\scriptsize,
	numberblanklines=true,
	showspaces=false,
	breaklines=true,
	showtabs=false,
	emph =
	{[1]
		class, package, interface, prototype
	},
	emphstyle={[1]\color{purple}\textbf},
	emph =
	{[2]
		internal, public, protected, private,
		super, this, import, new, extends, implements,
		void, true, false, as
	},
	emphstyle={[2]\color{flexblue}\textbf},
	emph =
	{[3]
		function
	},
	emphstyle={[3]\color{flexfunction}\textbf},
	emph =
	{[4]
		var
	},
	emphstyle={[4]\color{flexvar}\textbf}
}

    \lstdefinelanguage{ASN1}{%
      morekeywords=[1]{SEQUENCE,INTEGER,OPTIONAL,NumericString,IA5String,FALSE,ENUMERATED},
      morekeywords=[3]{},%
      morekeywords=[4]{},%
      morestring=[b]",%
     alsoletter={},
     alsoother={},
     sensitive%
    }[keywords,comments,strings]

\lstdefinelanguage{XHTML5}[]{HTML}{%
      morekeywords=[1]{mark,header,hgroup,nav,section,article},
      deletekeywords=[1]{width,height},
      morekeywords=[2]{width,height,svg:svg,svg:circle,cx,r,stroke,fill,xmlns:svg},
      alsoletter={:},
	morestring=[b]',
    }[keywords,comments,strings]

\lstdefinelanguage{pprolog}{%
  morestring=[b]",%
  literate=*{:-}{$\Leftarrow\;\,\,$}{2} {,\ }{$\ \land\ $}{2}
    {not}{$\neg$}{3} {!=}{$\neq$}{2}, 
}[keywords,strings]

\lstdefinelanguage{OPAL}[]{Prolog}{%
  morekeywords=[1]{TEMPLATE,INSTANTIATE,using},
  morekeywords=[2]{child,descendant,adjacent,following},
  morekeywords=[3]{form,link},
  morekeywords=[4]{concept,segment,unique},
  morestring=[b]",%
  alsoother={@},
  sensitive,
  literate=* {:-}{$\Leftarrow\;\,\,$}{2} {not\ }{$\neg$}{4} {and}{$\ \land\ $}{3}  
  {all}{$\forall\:$}{3} {\ or\ }{$\lor$}{4}  {!}{$\neg$}{1} 
  {!=}{$\neq$}{2} {...}{$\ldots$}{3} 
}[keywords,strings]

\lstdefinestyle{nonumbers}{numbers=none}
\lstdefinestyle{smaller}{basicstyle=\normalfont\ttfamily\footnotesize}
\lstdefinestyle{OPAL}{language=OPAL, numbers=none, mathescape=true,breaklines=false
  emphstyle=[1]{\normalfont\sffamily\scshape},
  emphstyle=[2]{\normalfont\sffamily\itshape}}

\usepackage[scaled=.8]{beramono}
\usepackage{paralist}
\lstnewenvironment{prolog}{\lstset{language=pprolog}}{}

\makeatletter
\newif\if@restonecol
\makeatother

\usepackage[ruled,vlined,linesnumbered,shortend]{algorithm2e}

\SetKwData{Cand}{CandDAs}
\SetKwData{Last}{LastDA}
\SetKwData{Curr}{CurrDA}
\SetKwData{Final}{FinalDAs}
\SetKwData{Depth}{Depth}
\SetKwData{Dist}{Dist}
\SetKwData{Nodes}{Nodes}
\SetKwFunction{pathLengths}{pathLengths}
\SetKwFunction{trimPrefix}{trimPrefix}
\SetKwData{lca}{lca}
\SetKwData{lspace}{lspace}
\SetKwData{StartCands}{StartCandidates}
\SetKw{Or}{or}
\SetKw{LogicalAnd}{and} 
\SetKw{Select}{select}
\SetKw{With}{with}
\SetKwInOut{Input}{input}
\SetKwInOut{Output}{output}
\SetKwInOut{Modifies}{modifies}
\SetKwData{Part}{Partition}
\SetKw{Add}{add}
\SetKw{Remove}{remove}

\makeatother

\usepackage{mathptmx}
\usepackage[scaled=.85]{helvet}
\usepackage{balance}

\begin{document}




\newcommand{\AckGiorgio}{Giorgio Orsi has also been supported by the Oxford Martin School's grant no. LC0910-019.}
\newcommand{\AckStandard}{The research leading to these results has received
    funding from the European Research Council under the European Community's
    Seventh Framework Programme (FP7/2007--2013) / ERC grant agreement DIADEM,
    no.~246858. }
\newcommand{\ACKNOWLEDGEMENTS}{\AckStandard\AckGiorgio}

\title{AMBER: Automatic Supervision for Multi-Attribute Extraction%
%
}


\author{Tim Furche \and Georg Gottlob \and Giovanni Grasso \and\\
  Giorgio Orsi \and Christian Schallhart \and Cheng Wang}
\institute{Oxford University Department of Computer Science\\
           Wolfson Building, Parks Road, Oxford OX1 3QD\\
          \email{firstname.lastname@cs.ox.ac.uk}}

\journalname{}
\date{25 Sep 2012}

\maketitle

\begin{abstract}
  The extraction of multi-attribute objects from the deep web is the bridge
  between the unstructured web and structured data.
  Existing approaches either induce wrappers from a set of human-annotated pages
  or leverage repeated structures on the page without supervision.  What the
  former lack in automation, the latter lack in accuracy. Thus accurate,
  automatic multi-attribute object extraction has remained an open challenge.
  
  \AMBER overcomes both limitations through mutual supervision between the
  repeated structure and automatically produced annotations. Previous approaches
  based on automatic annotations have suffered from low quality due to the
  inherent noise in the annotations and have attempted to compensate by
  exploring multiple candidate wrappers. In contrast, \AMBER compensates for
  this noise by integrating repeated structure analysis with annotation-based
  induction:
  The repeated structure limits the search space for wrapper
  induction, and conversely, annotations allow the repeated structure analysis
  to distinguish noise from relevant data. 
  Both, low recall and low precision in the annotations are mitigated to achieve
  almost human quality ($>98\%$) multi-attribute object extraction.  
  
  To achieve this accuracy, \AMBER needs to be trained once for an entire
  domain. \AMBER bootstraps its training from a small, possibly noisy set of
  attribute instances and a few unannotated sites of the domain.
\end{abstract}

\section{Introduction}
\label{sec:introduction}

The ``web of data'' has become a meme when talking about the future of the
web. Yet most of the objects published on the web today are only published
through HTML interfaces. Though structured data is increasingly available for
common sense knowledge such as Wikipedia, transient data such as product offers
is at best  available from large on-line shops such as Amazon or large-scale
aggregators.

The aim to extract objects together with their attributes from the web is
almost as old as the web. Its realisation has focused on exploiting two
observations about multi-attribute objects on the web:
\begin{inparaenum}[\bfseries(1)]
\item Such objects are typically presented as list, tables, grids, or other
  \emph{repeated structures} with a common template used for all objects.
\item Websites are designed for humans to quickly identify the objects and
  their attributes and thus use a \emph{limited visual and textual vocabulary}
  to present objects of the same domain. For example, most product offers contain a
  prominent price and image. 
\end{inparaenum}

Previous approaches have focused either on highly accurate, but supervised
extraction, where humans have to annotate a number of example pages for each
site, or on unsupervised, but low accuracy extraction based on detecting
repeated structures on any web page: \emph{Wrapper
induction}~\cite{DBLP:conf/sigmod/DalviBS09,%
  freitag00:_machin_learn_for_infor_exrtr,%
  hsu98:_gener_finit_state_trans_for,DBLP:journals/dke/KosalaBBB06,%
  kushmerick97:_wrapp_induc_for_infor_extrac,%
  muslea01:_hierar_wrapp_induc_for_semis_infor_system,%
  DBLP:conf/icde/GulhaneMMRRSSTT11} and semi-supervised
approaches~\cite{Baumgartner2001VisualWebIEwithLixto,%
  laender02:_debye} are of the first kind and require manually annotated
examples to generate an extraction program (\emph{wrapper}). Though such
annotations are easy to produce due to the above observations, it is
nevertheless a significant effort, as most sites use several types or variations
of templates that each need to be annotated separately: Even a modern wrapper
induction approach~\cite{DBLP:conf/icde/GulhaneMMRRSSTT11} requires more than 20
pages per site, as most sites require training for more than 10 different
templates.  Also, wrapper induction approaches are often focused on extracting a
single attribute instead of complete records, as for example
in~\cite{kushmerick97:_wrapp_induc_for_infor_extrac,DBLP:conf/sigmod/DalviBS09}.

On the other hand, the latter, fully \emph{unsupervised, domain-independent
approaches}~\cite{crescenzi02:_roadr,kayed10:_fivat,%
  liu03:_minin_data_recor_in_web_pages,%
  liu06:_vision_based_web_data_recor_extrac,simon05:_viper,%
  zhai06:_struc_data_extrac_from_web}, suffer from a lack of guidance on which
parts of a web site contain relevant objects: They often recognise irrelevant,
but regular parts of a page in addition to the actual objects and are
susceptible to noise in the regular structure, such as injected ads. Together
this leads to low accuracy even for the most recent approaches. This limits
their applicability for turning an HTML site into a structured database, but
fits well with web-scale extraction for search engines and similar settings,
where coverage rather than recall is essential (see
\cite{DBLP:journals/cacm/CafarellaHM11}): From every site some objects or pages
should be extracted, but perfect recall is not achievable at any rate and also
not necessarily desirable. To improve precision these approaches only consider
object extraction from certain structures, e.g.,
\emph{tables}~\cite{DBLP:journals/pvldb/CafarellaHWWZ08} or
\emph{lists}~\cite{DBLP:journals/vldb/ElmeleegyMH11}, and are thus not
applicable for general multi-attribute object extraction.

This lack of accurate, automated multi-attribute extraction has led to a recent
focus in data extraction approaches
\cite{DBLP:journals/pvldb/DalviKS11,SMM*08,DBLP:conf/icde/DerouicheCA12} on
coupling repeated structure analysis, exploiting observation \textbf{(1)}, with
automated annotations (exploiting observation \textbf{(2)}, that most websites
use similar notation for the same type of information). What makes this coupling
challenging is that both the repeated structure of a page and the automatic
annotations produced by typical annotators exhibit considerable noise.
\cite{DBLP:journals/pvldb/DalviKS11} and \cite{SMM*08} address both types of
noise, but in separation. In \cite{SMM*08} this leads to very low accuracy, in
\cite{DBLP:journals/pvldb/DalviKS11} to the need to considerable many
alternative wrappers, which is feasible for single-attribute extraction but
becomes very expensive for multi-attribute object extraction where the space of
possible wrappers is considerably larger. \cite{DBLP:conf/icde/DerouicheCA12}
addresses noise in the annotations, but relies on a rigid notation of separators
between objects for its template discovery which limits the types of noise it
can address and results in low recall.

To address these limitations, \AMBER tightly integrates repeated structure
analysis with automated annotations, rather than relying on a shallow
coupling. Mutual supervision between template structure analysis and annotations
allows \AMBER to deal with significant noise in both the annotations and the
regular structure without considering large numbers of alternative wrappers, in
contrast to previous approaches. Efficient mutual supervision is enabled by a
novel insight based on observation \textbf{(2)} above: that in nearly all
product domains there are one or more \emph{regular attributes}, attributes that
appear in almost every record and are visually and textually distinct. The most
common example is \TYPE{price}, but also the \TYPE{make} of a car or the
\TYPE{publisher} of a book can serve as regular attribute. By providing this
extra bit of domain knowledge, \AMBER is able to efficiently extract
multi-attribute objects with near perfect accuracy even in presence of
significant noise in annotations and regular structure.

Guided by occurrences of such a regular attribute, \AMBER performs a fully
automated repeated structure analysis on the annotated DOM to identify objects
and their attributes based on the annotations. It separates wrong or irrelevant
annotations from ones that are likely attributes and infers missing attributes
from the template structure.

\AMBER's analysis follows the same overall structure of the repeated structure
analysis in unsupervised, domain-independent approaches:
\begin{inparaenum}[\bfseries(1)]
\item \emph{data area identification} where \AMBER separates areas
  with relevant data from noise, such as ads or navigation menus,
\item \emph{record segmentation} where \AMBER splits data areas into
  individual records, and
\item \emph{attribute alignment} where \AMBER identifies the
  attributes of each record.
\end{inparaenum}
But \emph{unlike} these approaches, the first two steps are based on occurrences
of a regular attribute such as \TYPE{price}: Only those parts of a page where such
occurrences appear with a certain regularity are considered for data areas,
eliminating most of the noise produced by previous unsupervised approaches, yet
allowing us to confidently deal with pages containing multiple data
areas. Within a data area, theses occurrences are used to guide the segmentation
of the records.  Also the final step, attribute alignment, differs notably from
the unsupervised approaches: It uses the annotations (now for all attribute
types) to find attributes that appear with sufficient regularity on this page,
compensating both for low recall and for low precision.

Specifically, \AMBER's main contributions are:
\begin{asparaenum}[\bfseries(1)]
\item \AMBER is the first \emph{multi-attribute object extraction} system that
  combines \emph{very high accuracy} ($>95\%$) with zero site-specific supervision.

\item \AMBER achieves this by \emph{tightly integrating repeated structure analysis with
    induction from automatic annotations:}
  In contrast to previous approaches, it integrates these two parts to deal with
  noise in both the annotations and the regular structure, yet avoids
  considering multiple alternative wrappers by guiding the template structure
  analysis through annotations for a regular attribute type given as part of the
  domain knowledge:
  \begin{inparaenum}[\bfseries(a)]
  \item \emph{Noise in the regular structure:} \AMBER separates \emph{data
      areas} which contain relevant objects from noise on the page (including
    other regular structures such as navigation lists) by clustering annotations
    of regular attribute types according to their depth and distance on the
    page (Section~\ref{sec:data-area-identification}). \AMBER separates
    \emph{records}, i.e., regular occurrences of relevant objects in a data
    area, from noise between records such as advertisements through a regularity
    condition on occurrences of regular attribute types in a data area
    (Section~\ref{subsec:segmentation}).
  \item \emph{Noise in the annotations:} Finally, \AMBER addresses such noise
    by exploiting the regularity of attributes in records,
    compensating for low recall by inventing new attributes with sufficient
    regularity in other records, and for low precision by dropping annotations
    with insufficient such regularity
    (Section~\ref{subsec:attr-reconciliation}). We show that \AMBER can tolerate
    significant noise and yet attain
    above $98\%$ accuracy, dealing with, e.g., 50 false positive
    locations per page on average (Section~\ref{sec:evaluation}). 
  \item \emph{Guidance}: The annotations of regular attributes are also
    exploited to guide the search for a suitable wrapper, allowing us to
    consider only a few, local alternatives in the record segmentation
    (Section~\ref{subsec:segmentation}), rather than many 
    wrappers, as necessary in \cite{DBLP:journals/pvldb/DalviKS11} (see
    Section~\ref{sec:related-work}). 
  \end{inparaenum}
\item To achieve such high accuracy, \AMBER requires a thin layer of
  \emph{domain knowledge} consisting of annotators for the attribute types in
  the domain and the identification of a regular attribute type. In
  Section~\ref{sec:learning}, we give a \emph{methodology} for minimising the
  effort needed to create this domain knowledge: From a few example instances
  (collected in a \emph{gazetteer}) for each attribute type and a few,
  unannotated result pages of the domain, \AMBER can automatically bootstap
  itself by verifying and extending the existing gazetteers. This exploits
  \AMBER's ability to extract some objects even with annotations that have very
  low accuracy (around $20\%$).  Only for regular attribute types a
  reasonably accurate annotator is needed from the beginning. This is easy to
  provide in product domains where price is such an attribute type. In other
  domains, we have found producers such as book publishers or car makers a
  suitable regular attribute type for which accurate annotators are also easy to
  provide.
\item 
  We \emph{evaluate} \AMBER on the UK
  real-estate and used cars markets against a gold standard consisting
  of \emph{manually} annotated pages from 150 real estate sites (281
  pages) and 100 used car sites (150 pages).
  Thereby, \AMBER is \emph{robust} against significant noise:
  Increasing the error rate in the annotations from $20\%$ to over
  $70\%$, drops \AMBER's accuracy by only $3\%$.
  (Section~\ref{sec:qual-eval}).
  \begin{inparaenum}[\bfseries(a)]
  \item We \emph{evaluate} \AMBER on 2,215 pages from 500 real estate
    sites by \emph{automatically} checking the number of extracted
    records (20,723 records) and related attributes against the
    expected extrapolated numbers (Section~\ref{sec:quan-eval}).
  \item We compare \AMBER with \ROADRUNNER~\cite{crescenzi02:_roadr}
    and \MDR~\cite{liu03:_minin_data_recor_in_web_pages},
    demonstrating \AMBER's superiority
    (Section~\ref{sec:comp-with-other}).
  \item At last, we show that \AMBER can learn a gazetteer from a seed
    gazetteer, containing 20\% of a complete gazetteer, thereby
    improving its accuracy from 50.5\% to 92.7\%.
  \end{inparaenum}
\end{asparaenum}

While inspired by earlier work on rule-driven result page
analysis~\cite{furche11:_littl_knowl_rules_web}, this paper is the first
complete description of \AMBER as a self-supervised system for extracting
multi-attribute objects. In particular, we have redesigned the integration
algorithm presented in Section~\ref{sec:approach} to deal with noise in both
annotators and template structure. We have also reduced the amount of domain
knowledge necessary for \AMBER and provide a methodology for semi-supervised
acquisition of that domain knowledge from a minimal set of examples, once for an
entire domain. Finally, we have significantly expanded the evaluation to reflect
these changes, but also to provide deeper insight into \AMBER.

\subsection{Running Example}
\label{sec:running-example}

\begin{figure}[tbp]
  \centering
  \includegraphics[width=\linewidth]{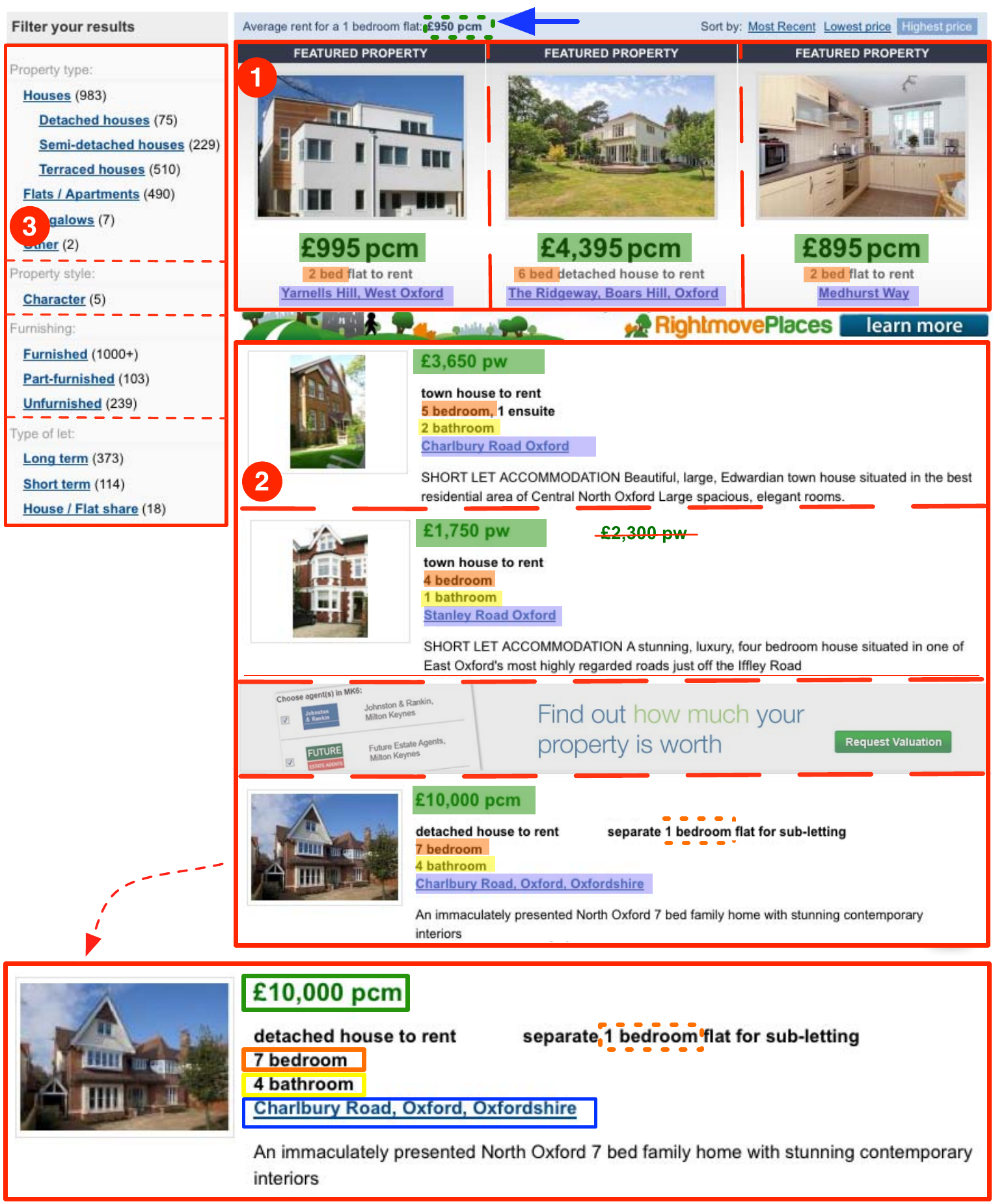}
  \caption{Result Page on \texttt{rightmove.co.uk} }
 \label{fig:rightmove-example}
\end{figure}

We illustrate \AMBER on the result page from Rightmove,
the biggest UK real estate aggregator. Figure~\ref{fig:rightmove-example} shows
the typical parts of such pages: 
On top,
\begin{inparaenum}[\bfseries(1)]
\item some featured properties are arranged in a horizontal block,
  while directly below, separated by an advertisement,
\item the properties matching the user's query are listed vertically.
  Finally, on the left-hand side, a block
\item provides some filtering options to refine the search result.
\end{inparaenum}
At the bottom of Figure~\ref{fig:rightmove-example} we zoom into the
third record, highlighting the identified attributes.

After annotating the DOM of the page, \AMBER analyzes the page in
three steps: data area identification, record segmentation, and
attribute alignment.
In all these steps we exploit annotations provided by domain-specific
annotators, in particular for regular attribute types, here \TYPE{price}, to
distinguish between relevant nodes and noise such as ads.
%

For Figure~\ref{fig:rightmove-example}, \AMBER identifies \type{price}
annotations (highlighted in green, e.g., ``\pounds 995 pcm''), most locations
(purple), the number of bedrooms (orange) and bathrooms (yellow).
The price on top (with the blue arrow), the ``1 bedroom'' in the third record,
and the crossed out price in the second record are three examples of false
positives annotations, which are corrected by \AMBER subsequently.

\paragraph{Data area identification.} First, \AMBER detects which
parts of the page contain relevant data.
In contrast to most other approaches, \AMBER deals with web pages
displaying multiple, differently structured data areas.
E.g., in Figure~\ref{fig:rightmove-example} \AMBER identifies two data
areas, one for the horizontally repeated featured properties and one
for the vertically repeated normal results (marked by red boxes).

Where other approaches rely solely on repeated structure
, \AMBER first identifies \emph{pivot nodes,} i.e.,
nodes on the page that contain annotations for regular attribute types, here
\type{price}.
Second, \AMBER obtains the data areas as clusters of continuous
sequences of pivot nodes which are evenly spaced at roughly
the same DOM tree depth and distance from each other.
For example, \AMBER does not mistake the filter list (3) as a data area,
despite its large size and regular structure.
Approaches only analyzing structural or visual structures may fail to
discard this section. 
Also, any annotation appearing outside the found areas is discarded,
such as the price annotation with the blue arrow atop of area (1).

\paragraph{Record segmentation.} Second, \AMBER needs to segment the
data area into ``records'', each representing one multi-attribute
object.
To this end, \AMBER cuts off noisy pivot nodes at the head and tail of
the identified sequences and removes interspersed nodes, such as the
crossed out price in the second record.
The remaining pivot nodes segment the data area into fragments of uniform size,
each with a highly regular structure, but additional shifting may be required as
the pivot node does not necessarily appear at the beginning of the record. Among
the possible record segmentations the one with highest regularity among the
records is chosen.
In our example, \AMBER correctly determines the records for the data
areas (1) and (2), as illustrated by the dashed lines.
\AMBER prunes the advertisement in area (2) as inter-record noise, since it
would lower the segmentation regularity.

\paragraph{Attribute alignment.} Finally, \AMBER aligns the found
annotations within the repeated structure to identify the record
attributes.
Thereby, \AMBER requires that each attribute occurs in sufficiently
many records at corresponding positions.
If this is the case, it is \emph{well-supported,} and otherwise, the
annotation is dropped.
Conversely, a missing attribute is inferred, if sufficiently many
records feature an annotation of the same type at the position in
concern.
For example, all location annotations in data area $2$ share the same
position, and thus need no adjustment.
However, for the featured properties, the annotators may fail to recognize
``Medhurst Way'' as a location.
\AMBER infers nevertheless that ``Medhurst Way'' must be a location
(as shown in Figure~\ref{fig:rightmove-example}), since all other records 
have a location at the corresponding position.
For data area $2$, bathroom and bedroom number are shown respectively
at the same relative positions.
However, the third record also states that there is a separate flat
to sublet with one bedroom.
This node is annotated as bedroom number, but \AMBER recognizes it is
false positive due to the lack of support from other records.

\bigskip
To summarise, \AMBER addresses low recall and precision of annotations in the
attribute alignment, as it can rely on an already established record
segmentation to determine the regularity of the attributes. In addition it
compensates for noise in the annotations for regular attribute types in the
record segmentation by majority voting to determine the length of a record and
by dropping irregular annotations (such as the crossed out price in record
2). \AMBER also addresses noise in the regular structure on the page, such as
advertisements between records and regular, but irrelevant areas on the page
such as the refinement links. All this comes at the price of requiring some
domain knowledge about the attributes and their instances in the domain, that
can be easily acquired from just a few examples, as discussed in
Section~\ref{sec:learning}.

\section{Multi-Attribute Object Extraction}
\label{sec:approach-def}

\subsection{Result Page Anatomy}
\label{sec:result-pages}

\AMBER extracts multi-attribute objects from \emph{result pages},
i.e., pages that are returned as a response to a form query on a web
site.
The typical anatomy of a result page is a repeated structure of more or less
complex records, often in form of a simple sequence. 
Figure~\ref{fig:rightmove-example} shows a typical case, presenting a
paginated sequence of records, each representing a real estate property to rent,
with a price, a location, the number of bed and bath rooms.
%

We call \emph{record} each instance of an object on the page and we
refer to a group of continuous and similarly structured records as
\emph{data area.}
Then, result pages for a schema $\SCHEMA=\REGATTS\cup\OPTATTS$ that defines the
optional and regular attribute types of a domain have the following
characteristics:
\begin{inparaenum}[\bfseries(D1)]
  Each data area consists of 
%
%
\item a \emph{maximal} and
\item \emph{continuous sequence of records,} while each record
\item is a sequence of children of the data area root, and 
  consists of
\end{inparaenum}
\begin{inparaenum}[\bfseries(R1)]
\item a \emph{continuous sequence of sibling subtrees} in the DOM
  tree.
  For all records, this sequence is of
\item the \emph{same length,} of
\item the \emph{same repeating structure,} and contains
\item in most cases \emph{one} instance of \emph{each regular
    attribute} in \REGATTS.
  Furthermore, each record may contain
\item instances of some \emph{optional attributes} \OPTATTS, such that
  attributes for all attribute types in $\REGATTS\cup\OPTATTS$
\item appear at \emph{similar positions} within each record, if they appear at all.
\end{inparaenum}
For attributes, we note that relevant attributes 
\begin{inparaenum}
\item[\bfseries(A1)] tend to appear early within their record, with 
\item[\bfseries(A2)] its textual content filling a large part of their surrounding
  text box. Also
\item[\bfseries(A3)] attributes for optional attribute types tend to be less
  standardized in their values, represented with more variations.
\end{inparaenum}

\begin{figure}[tbp]
  \centering
  \includegraphics[width=.85\columnwidth]{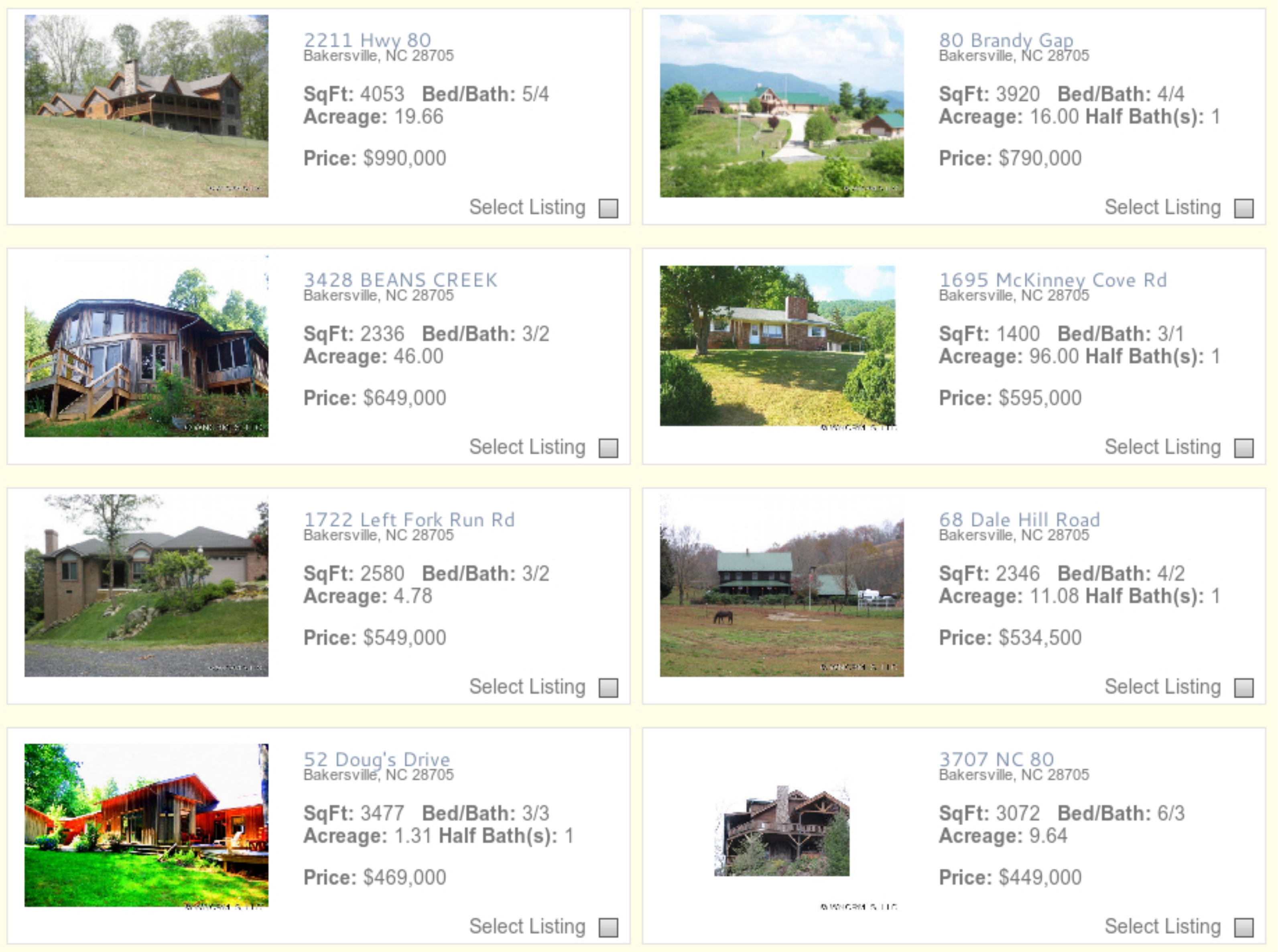}
  \caption{A grid result page.}
  \label{fig:rp-grid}
\end{figure}

Result pages comes in many shapes, e.g., \emph{grids}, like the one depicted in
Figure~\ref{fig:rp-grid} taken from the \url{appalachianrealty.com} real estate
website, \emph{tables}, or even simple \emph{lists}. The prevalent case,
however, is the sequence of individual records as in Figure~\ref{fig:rightmove-example}.

Many result pages on the web are regular, but many also contain considerable
noise.
In particular, an analysis must 
\begin{inparaenum}[\bfseries(N1)]
\item tolerate inter-record noise, such as advertisements between records, and
\item intra-record noise, such as instances of attribute types such as
  \TYPE{price} occurring also in product descriptions. 
  It must also 
\item address pages with multiple data areas distinguish them from regular, but
  irrelevant noise. .
\end{inparaenum}

\begin{figure}[tbp]
  \centering
  \includegraphics[width=.85\columnwidth]{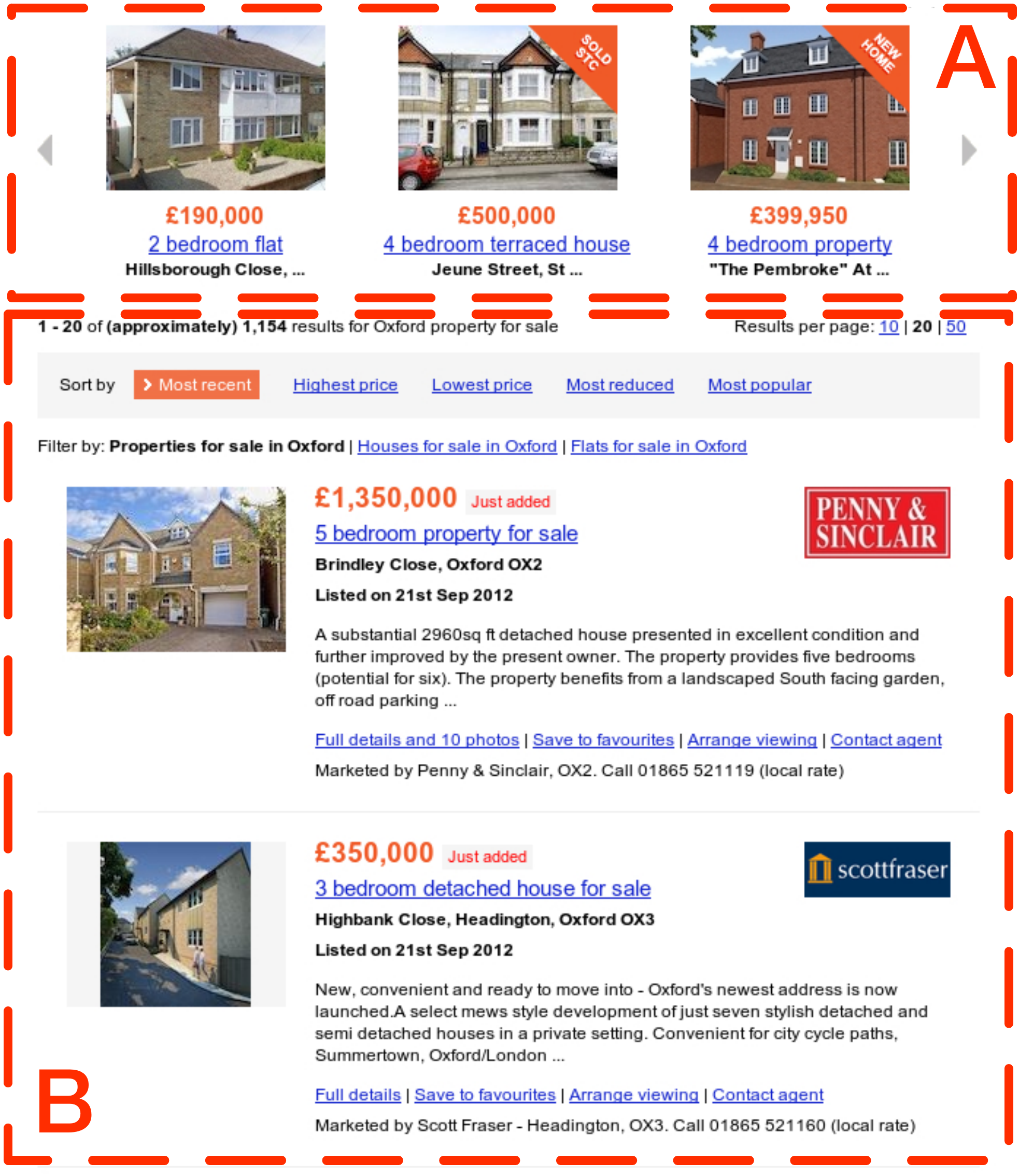}
  \caption{Multiple data areas.}
  \label{fig:rp-multiple-dataarea}
\end{figure}

\paragraph*{Further Examples.}\ 
Consider a typical result page from \url{Zoopla.co.uk}
(Figure~\ref{fig:rp-multiple-dataarea}). 
Here we have two distinct data areas where records are laid out using
different templates.
Premium (i.e., sponsored) results appear in the top data area {\bf
  (A)}, while regular results appear in the bottom data area {\bf
  (B)}.
A wrapper generation system must be able to cluster the two kinds of
records and distinguish the different data areas.
Once the two data areas have been identified, the analysis of the
records does not pose particular difficulties since, within each data
area, the record structure is very regular.

\begin{figure}[tbp]
  \centering
  \includegraphics[width=.85\columnwidth]{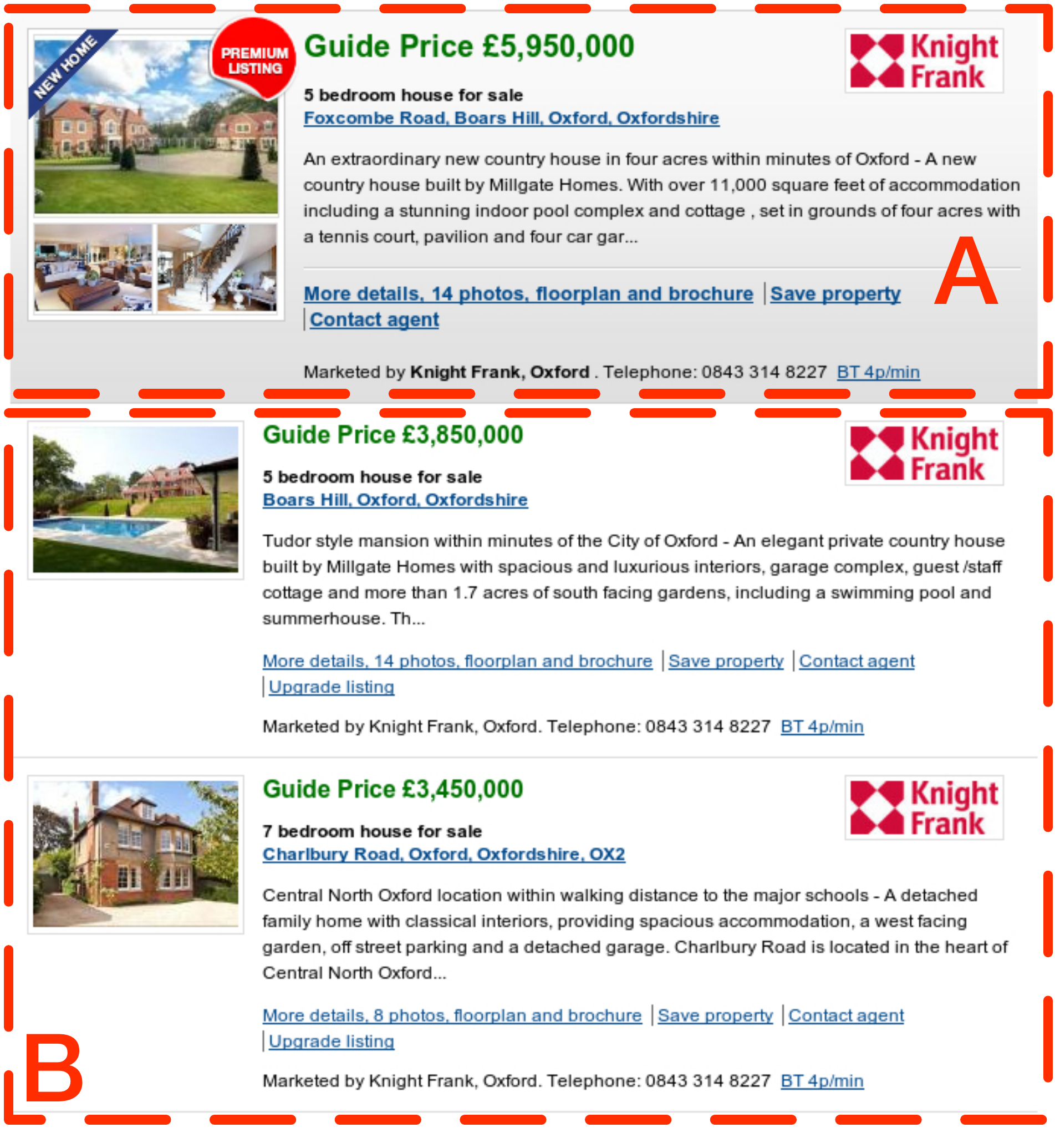}
  \caption{Premium records}
  \label{fig:rp-premium}
\end{figure}

Another interesting case is the presence of highlighted results like
in Figure~\ref{fig:rp-premium}, again taken from
\url{Rightmove.co.uk}, where premium records {\bf (A)} are diversified
from other results {\bf (B)} within the same data area.
This form of highlighting can easily complicate the analysis of the
page and the generation of a suitable wrapper.



\subsection{Extraction Typing}
\label{sec:extraction-typing}

For extracting multi-attribute objects, we output a data structure describing
each object and its attributes, such as origin, departure time, and price.
In addition, to automatically induce wrappers, \AMBER needs not only to extract
this data but must also link the extracted data to its representation on the
originating pages.
To that end, \AMBER types nodes in the DOM for extraction (\emph{extraction
  typing}) to describe
\begin{inparaenum}[\bfseries(1)]
\item how objects appear on the page as \emph{records},
\item how \emph{attributes} are structured within records,
  and
\item how records are grouped into \emph{data areas.}
\end{inparaenum}
In supervised wrapper induction systems, this typing is usually
provided by humans ``knowing'' the objects and their attributes. But in fully
unsupervised induction, also the generation of the extraction typing is
automated.
To formalise extraction typing, we first define a web page and then type its
nodes according to a suitable domain schema.

%

\paragraph*{Web pages.}\ 
Following~\cite{Benedikt2007XPath-Leashed}, we represent a web page
as its DOM tree $P = \bigl((U)_{U\in\unary}, \DOMREL{child},
\DOMREL{next-sibl}\bigr)$ where each $\lambda \in (U)_{U \in \unary}$
is a unary relation to label nodes with $\lambda$,
$\DOMREL{child}(p,c)$ holds if $p$ is a parent node of $c$, and
$\DOMREL{next-sibl}(s,s')$ holds if $s'$ is the sibling directly
following $s$.
In abuse of notation, we refer to $P$ also as the set of DOM nodes in
$P$. 
Further relations, e.g., \DOMREL{descendant} and \DOMREL{following},
are derived from these basic relations.
We write $x \PRECEDES y$, if $x$ is a preceding sibling of $y$, and we
write $x\PRECEDESEQ y$ for $x\PRECEDES y$ or $x=y$.
For all nodes $n$ and $n'$, we define the sibling distance $n \SDIST
n'$ with 
$$
n \SDIST n'=\left\{
\begin{array}{lcl}
  n\PRECEDESEQ n' & : & \hfill |\{ k \mid n\PRECEDES k \PRECEDESEQ n'\}| \\
  n'\PRECEDES n & : & -|\{ k \mid n'\PRECEDES k \PRECEDESEQ n\}| \\
  \text{otherwise} & : & \infty
\end{array}\right.
$$
Finally, $\DOMREL{first-child}(p,c)$ holds if $c$ is the first child
of $p$, i.e., if there is no other child $c'$ of $p$ with $c'
\PRECEDES c$.

\paragraph*{Extraction Typing.}\ Intuitively, data areas, records,
and attributes are represented by (groups of) DOM nodes.
An extraction typing formalizes this in typing the nodes accordingly
to guide the induction of a suitable wrapper for pages generated from
the same template and relies on a domain schema for providing attribute
types. We distinguish attribute types into regular and optional, the latter
indicating that attributes of that type typically occur only in some, but not
all records. 

\begin{compactdef}
  A \textbf{domain schema} $\SCHEMA = \REGATTS \cup \OPTATTS$ defines
  disjoint sets $\REGATTS$ and $\OPTATTS$ of regular and optional
  attribute types.
\end{compactdef}

\begin{compactdef}
  \label{def:extraction-typing}
  Given a web page with DOM tree $P$, an \textbf{extraction typing}
  for domain schema $\SCHEMA=\REGATTS \cup \OPTATTS$ is a relation
  $\TYPING: P \times (\SCHEMA \cup \{\dataareatype, \recordstarttype,
  \recordtailtype\})$ where each node $n\in P$ with
  \begin{compactenum}[\bfseries(1)]
  \item $\TYPING(n,\dataareatype)$ contains a data area, with 
  \item $\TYPING(n,\recordstarttype)$, $n$ represents a record that
    spans the subtrees rooted at $n$ and its subsequent siblings $n'$.
    For all these subsequent siblings $n'$, we have
  \item $\TYPING(n',\recordtailtype)$, marking the tail of the record.
  \item $\TYPING(n, \rho)$ holds, if $n$ contains an attribute of type
    $\rho\in\SCHEMA$.
  \end{compactenum}
  Data areas may not be nested, neither may records, but records must be
  children of a data area, and attributes must be descendants of a
  (single) record.
\end{compactdef}


\begin{compactdef}
  \label{def:partof}
  Given an extraction typing \TYPING, a node $n$ \textbf{is part of} a
  record $r$, written $\partOf_\TYPING(n,r)$, if the following
  conditions hold:
  \begin{inparaenum}[]
  \item $\TYPING(r,\recordstarttype)$ holds, 
  \item $n$ occurs in a subtree rooted at node $r'$ with
    $\TYPING(r',\recordstarttype)$ or $\TYPING(r',\recordtailtype)$,
    and
  \item there is no node $r''$ between $r$ and $r'$ with
    $\TYPING(r'',\recordstarttype)$.
  \end{inparaenum}
  A record $r$ is part of a data area $d$, written
  $\partOf_\TYPING(r,d)$, if $r$ is a child of $d$, and transitively,
  we have $\partOf_\TYPING(n,d)$ for 
  $\partOf_\TYPING(n,r)$ and $\partOf_\TYPING(r,d)$.
\end{compactdef}

\begin{figure*}[tbp]
  \centering
  \includegraphics[width=.9\linewidth]{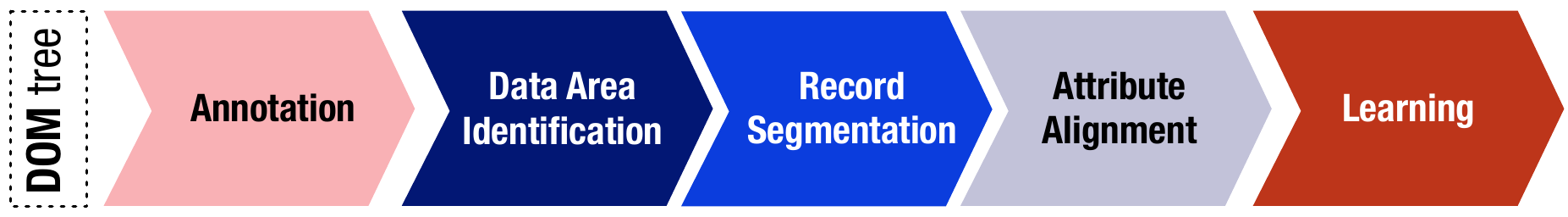}
  \caption{\AMBER workflow}
  \label{fig:workflow}
\end{figure*}

\section{The \AMBERnofont Approach}
\label{sec:approach}

Following the result page anatomy from the preceding section, the
extraction of multi-attribute objects involves three main tasks:
\begin{inparaenum}[\bfseries(1)]
\item \emph{Identifying data areas} with relevant objects among other
  noisy contents, such as advertisements or navigation menus,
\item \emph{segmenting} such data areas into \emph{records}, i.e.,
  representations of individual objects, and
\item \emph{aligning attributes} to objects, such that all records
  within the same data area feature a similar attribute structure.
\end{inparaenum}

An attempt to exploit properties {\bfseries(D1-3)}, {\bfseries(R1-6)},
and {\bfseries(A1-3)} directly, leads to a circular search:
Data areas are groups of regularly structured records, while records
are data area fragments that exhibit structural similarities with all
other records in the same area.
Likewise, records and attributes are recognized in mutual reference to
each other.
Worse, automatically identifying attribute values is a naturally noisy
process based on named entity recognition (e.g., for locations) or
regular expressions (e.g., for postcodes or prices).
Hence, to break these cyclic dependencies, we draw some basic
consequences from the above characterization.
Intuitively, these properties ensure that the instances of each
regular attribute $\rho\in\REGATTS$ constitute a cluster in each data
area,
\begin{inparaenum}[\bfseries(D1)]\setcounter{enumi}{3}
  where each instance occurs
\item roughly at the \emph{same depth} in the DOM tree and
\item roughly at the \emph{same distance.}
\end{inparaenum}

Capitalizing on these properties, and observing that it is usually
quite easy to identify the regular attributes $\REGATTS$ for specific
application domains, \AMBER relies on occurrences of those regular
attributes to determine the records on a page:
Given an annotator for a single such attribute $\pi\in\REGATTS$
(called \textbf{pivot attribute type}), \AMBER fully automatically
identifies relevant data areas and segments them into records.
Taking advantage of the repeating record structure, this works well,
even with fairly low quality annotators, as demonstrated in
Section~\ref{sec:evaluation}.
For attribute alignment, \AMBER requires corresponding annotators for
the other domain types, also working with low quality annotations.
For the sake of simplicity, we ran \AMBER with a single pivot
attribute per domain~-- achieving strong results on our evaluation
domains (UK real estate and used car markets).
However, one can run \AMBER in a loop to analyze each page
consecutively with different pivot attributes to choose the extraction
instance which covers most attributes on the page.

Once, a pivot attribute type has been chosen, \AMBER identifies and
segments data areas based on \textbf{pivot nodes,} i.e., DOM nodes
containing instances of the pivot attribute:
Data areas are DOM fragments containing a cluster of pivot nodes
satisfying {\bfseries(D4)} and {\bfseries(D5)}, and records are
fragments of data areas containing pivot nodes in similar positions.
Once data areas and records are fixed, we refine the attributes
identified so far by aligning them across different records and adding
references to the domain schema.
With this approach, \AMBER deals incomplete and noisy annotator (see
Section~\ref{sec:learning}), created with little effort, but still
extracts multi-attribute objects without significant overhead, as
compared to single attribute extraction.

Moreover, \AMBER deals successfully with the noise occurring on pages,
i.e., it
\begin{inparaenum}[\bfseries(N1)]
\item tolerates inter-record noise by recognizing the relevant data
  via annotations, 
\item tolerates intra-record variances by segmenting records driven by
  regular attributes,
  and it
\item address multi-template pages by considering each data area
  separately for record segmentation.
\end{inparaenum}

\subsection{Algorithm Overview}
\label{sec:algorithm-overview}

The main algorithm of \AMBER, shown in Algorithm~\ref{algo:amber} and
Figure~\ref{fig:workflow}, 
takes as inputs a DOM tree $P$ and a schema
$\SCHEMA=\REGATTS\cup\OPTATTS$, with a regular attribute type
$\pi\in\REGATTS$ marked as pivot attribute type, to produce an
extraction typing \TYPING. 
First, the annotations \ANNOT for the DOM $P$ are computed as
described in Section~\ref{sec:annotation-model}
(Line~\ref{algo:amb:annot}).
Then, the extraction typing \TYPING is constructed in three steps, by
identifying and adding the data areas (Line~\ref{algo:amb:da}), then
segmenting and adding the records (Line~\ref{algo:amb:rs}), and
finally aligning and adding the attributes (Line~\ref{algo:amb:aa}).
All three steps are discussed in
Sections~\ref{sec:data-area-identification}
to~\ref{subsec:attr-reconciliation}.
Each step takes as input the DOM $P$ and the incrementally expanded
extraction typing $\TYPING$.
The data area identification takes as further input the pivot
attribute type $\pi$ (but not the entire schema \SCHEMA), together
with the annotations \ANNOT. 
It produces~-- aside the data areas in \TYPING~-- the sets
$\PIVOTS(d)$ of pivot nodes supporting the found data areas $d$.
The record segmentation requires these \PIVOTS to determine the record
boundaries to be added to \TYPING, working independently from \SCHEMA.
Only the attribute alignment needs the schema \SCHEMA to type the DOM
nodes accordingly.
At last, deviances between the extraction typing \TYPING and the
original annotations \ANNOT are exploited in improving the gazetteers
(Line~\ref{algo:amb:learn})~-- discussed in Section~\ref{sec:learning}. 

\begin{algorithm}[tbp]\small
\SetKw{new}{new}
\SetKw{LET}{let}
\SetKwFunction{add}{add}
\SetKwFunction{size}{size}
\SetKwFunction{remove}{remove}
\SetKwFunction{maxPath}{maxPath}
\SetKwData{Support}{Support}
\SetKwData{maxDepth}{maxDepth}
\SetKwData{DistInt}{DistInt}
\SetKwData{minDist}{minDist}
\SetKwData{maxDist}{maxDist}
\SetKwData{Diff}{Diff}
\SetKwData{DiffDepth}{DiffDepth}
\SetKwData{DiffDist}{DiffDist}

\Input{$P$ -- DOM to be analyzed}
\Input{$\SCHEMA=\REGATTS\cup\OPTATTS$\ -- schema for the searched
  results, with a specifically marked pivot attribute $\pi\in\REGATTS$}
\Output{$\TYPING$ -- extraction typing on $P$}

\annotate$(P,\SCHEMA, \ANNOT)$\;\label{algo:amb:annot}
\dataareaalgo$(P, \TYPING, \pi, \ANNOT,\PIVOTS)$\;\label{algo:amb:da}
\recordalgo$(P, \TYPING, \PIVOTS)$\;\label{algo:amb:rs}
\attributealgo$(P,\TYPING,\Sigma)$\;\label{algo:amb:aa}
\learn(\TYPING,\ANNOT)\;\label{algo:amb:learn}

\caption{$\amberalgo(P, \TYPING, \SCHEMA)$}
\label{algo:amber}
\end{algorithm}

\subsection{Annotation Model}
\label{sec:annotation-model}

During its first processing step, \AMBER annotates a given input DOM
to mark instances of the attribute types occurring in
$\SCHEMA=\REGATTS\cup\OPTATTS$.
We define these annotations with a relation $\ANNOT: \SCHEMA \times
\DOMNODES \times \UNIVERSE$, where $\DOMNODES$ is the DOM node set,
and $\UNIVERSE$ is the union of the domains of all attribute types in
$\SCHEMA$.
$\ANNOT(A, n, v)$ holds, if $n$ is a text node containing a
representation of a value $v$ of attribute type $A$.
For the HTML fragment \lstinline|<span>Oxford,$\pounds$2k</span>|, we
obtain, e.g., $\ANNOT(\ATYPE{location}, t, \textrm{``Oxford''})$ and
$\ANNOT(\ATYPE{price}, t, \textrm{``2000''})$, where $t$ is the text
node within the \texttt{span}.

In \AMBER, we implement $\ANNOT$ with GATE, relying on a mixture of
manually crafted and automatically extracted gazetteers, taken from
sources such as DBPedia~\cite{Auer07dbpedia:a}, along with regular
expressions for prices, postcodes, etc.
In Section~\ref{sec:evaluation}, we show that \AMBER easily
compensates even for very low quality annotators, thus requiring only
little effort in creating these annotators.

\subsection{Data Area Identification}
\label{sec:data-area-identification}

We overcome the mutual dependency of data area, record, and attribute
in approximating the regular record through instances of the pivot
attribute type $\pi$:
For each record, we aim to identify a single \emph{pivot node}
containing that record attribute $\pi$ {\bfseries(R4)}. 
A data area is then a cluster of pivot nodes appearing regularly,
i.e., the nodes occur have roughly the same depth {\bfseries(D4)} and
a pairwise similar distance {\bfseries(D5)}.

Let $N_\pi$ be a set of pivot nodes, i.e., for each $n \in N_\pi$
there is some $v$ such that $\ANNOT(\pi,n,v)$ holds. 
Then we turn properties {\bfseries(D4)} and {\bfseries(D5)} into two
corresponding regularity measures for $N_\pi$:
\begin{inparaenum}[\bfseries(M1)]\setcounter{enumi}{3}
  $N_\pi$ is 
\item \emph{$\DEPTHTHRES$-depth consistent,} if there exists a $k$
  such that $\DEPTH(n) = k \pm \DEPTHTHRES$ for all $n \in N_\pi$,
  and $N_\pi$ is
\item \emph{$\DISTTHRES$-distance consistent,} if there
  exists a $k$ such that $|\textsf{path}(n,n')|=k \pm \DISTTHRES$ for
  all $n\neq n' \in N_\pi$. 
\end{inparaenum}
Therein, $\DEPTH(n)$ denotes the depth of $n$ in the DOM tree, and
$|\textsf{path}(n,n')|$ denotes the length of the undirected path from
$n$ to $n'$.
Assuming some parametrization $\DEPTHTHRES$ and $\DISTTHRES$, we
derive our definition of data areas from these measures:

\begin{compactdef}\label{def:data-area}
  A \textbf{data area} (for a regular attribute type $\pi$) is a
  maximal subtree $d$ in a DOM $P$ where
  \begin{compactenum}[\bfseries(1)]
  \item $d$ contains a set of pivot nodes $N_\pi$ with $|N_\pi| \geq
    2$,
  \item $N_\pi$ is depth and distance consistent {\bfseries (M4-5)},
  \item $N_\pi$ is maximal {\bfseries(D1)} and continuous {\bfseries
      (D2)}, and
  \item $d$ is rooted at the least common ancestor of $N_\pi$.
  \end{compactenum}
\end{compactdef}

\begin{algorithm}[tbp]\small
\SetKw{new}{new}
\SetKw{LET}{let}
\SetKwFunction{add}{add}
\SetKwFunction{size}{size}
\SetKwFunction{remove}{remove}
\SetKwFunction{maxPath}{maxPath}
\SetKwData{Support}{Support}
\SetKwData{maxDepth}{maxDepth}
\SetKwData{DistInt}{DistInt}
\SetKwData{minDist}{minDist}
\SetKwData{maxDist}{maxDist}
\SetKwData{Diff}{Diff}
\SetKwData{DiffDepth}{DiffDepth}
\SetKwData{DiffDist}{DiffDist}

\Input{$P$ -- DOM to be analyzed}
\Output{$\TYPING$ -- extraction typing on $P$ with data areas only}
\Input{$\pi$\ -- the pivot attribute type $\pi\in\REGATTS$}
\Input{\ANNOT\ --  annoations on $P$}
\Output{$\PIVOTS$ -- data areas support}

$\PIVOTS(n) \gets \emptyset$ for all $n \in P$\;\label{alg:da:initpivots}
$\Cand \gets  \{\;\;(\{n\}, [\DEPTH(n),\DEPTH(n)], [\infty,0])\;\;\mid\;\; \ANNOT(\pi,n,v)\}$\;\label{alg:da:candidateinit1}
$\Cand.add(\emptyset, [0,\infty],[0,\infty])$\;\label{alg:da:candidateinit2}
$\Last=(\Nodes_{\Last},\Depth_{\Last},\Dist_{\Last}) \gets (\emptyset, [], [])$\;\label{alg:da:initlastda}
\ForEach{$(\Nodes,\Depth,\Dist) \in \Cand$ in document order}{\label{alg:da:loopcandidates}
  $\Depth' \gets \Depth_{\Last} \intMerge \Depth$\;\label{alg:da:newdepth}
  $\Dist' \gets \Dist_{\Last} \intMerge \Dist \intMerge \pathLengths(\Nodes_{\Last}, \Nodes)$\;\label{alg:da:newdistance}
  \If{$|\Depth'| < \DEPTHTHRES$ \LogicalAnd $|\Dist'| < \DISTTHRES$}{\label{alg:da:depthdistancecheck}
    \tcc{Cluster can be extended further}
    $\Last \gets (\Nodes_{\Last} \cup \Nodes, \Depth', \Dist')$\;\label{alg:da:domerge}
  }
  \Else{
    \tcc{Cluster cannot be extended further}
    \If{$|\Nodes_{\Last}| \geq 2$}{\label{alg:da:sizecheck}
      $d \gets \lca(\Nodes_{\Last})$\;\label{alg:da:lca}
      \If{$|\PIVOTS(d)| < |\Nodes_{\Last}|$}{\label{alg:da:nolargerdacheck}
        $\PIVOTS(d) \gets \Nodes_{\Last}$; \Add $\TYPING(d,\dataareatype)$\;\label{alg:da:doadd}
      }
    }
    $\Last \gets (\Nodes,\Depth,\Dist)$\;\label{alg:da:lastdareset}
  }
}
\caption{$\dataareaalgo(P, \TYPING, \pi, \ANNOT, \PIVOTS)$}
\label{algo:data-area}
\end{algorithm}

Algorithm~\ref{algo:data-area} shows \AMBER's approach to identifying
data areas accordingly.
The algorithm takes as input a DOM tree $P$, an annotation relation
\ANNOT, and a pivot attribute type $\pi$. As a result, the algorithm
marks all data area roots $n\in P$ in adding
$\TYPING(n,\dataareatype)$ to the extraction typing \TYPING.
In addition, the algorithm computes the \emph{support} of each data
area, i.e., the set of pivot nodes giving rise to a data area.
The algorithm assigns this support set to $\PIVOTS(n)$, for use by the
the subsequent record segmentation.

The algorithm clusters pivot nodes in the document, recording for each
cluster the depth and distance interval of all nodes encountered so
far.
Let $I = [i_1,i_2]$ and $J = [j_1,j_2]$ be two such intervals. 
Then we define the merge of $I$ and $J$, $I \intMerge J =
[\min(i_1,j_1), \max(i_2,j_2)]$.
A (candidate) cluster is given as tuple $(\Nodes, \Depth, \Dist)$
where $\Nodes$ is the clustered pivot node set, and $\Depth$ and
$\Dist$ are the minimal intervals over $\mathbb{N}_0$, such that
$\DEPTH(n) \in \Depth$ and $|\textsf{path}(n,n')| \in \Dist$ holds for
all $n, n' \in \Nodes$.

During initialization, the algorithm resets the support $\PIVOTS(n)$
for all nodes $n\in P$ (Line~\ref{alg:da:initpivots}), turns all pivot
nodes into a candidate data areas of size 1
(Line~\ref{alg:da:candidateinit1}), and adds a special candidate data
area $(\emptyset,[0,\infty],[0,\infty])$
(Line~\ref{alg:da:candidateinit2}) to ensure proper termination of the
algorithm's main loop.
This data area is processed after all other data areas and hence
forces the algorithm in its last iteration into the else branch of
Line~\ref{alg:da:sizecheck} (explained below).
Before starting the main loop, the algorithm initializes
$\Last=(\Nodes_{\Last},\Depth_{\Last},\Dist_{\Last})$ to hold the data
area constructed in the last iteration.
This data area is initially empty and set to $(\emptyset, [], [])$
(Line~\ref{alg:da:initlastda}).

After initialization, the algorithm iterates in document order over
all candidate data areas $(\Nodes,\Depth,\Dist)$ in $\Cand$
(Line~\ref{alg:da:loopcandidates}).
In each iteration, the algorithm tries to merge this data area with
the one constructed up until the last iteration, i.e., with $\Last$.
If no further merge is possible, the resulting data area is added as a
result (if some further property holds).
To check whether a merge is possible, the algorithm first merges the
depth and distance intervals (Lines~\ref{alg:da:newdepth}
and~\ref{alg:da:newdistance}, respectively).
The latter is computed by merging the intervals from the clusters with
a third one, $\pathLengths$, the interval covering the path lengths
between pairs of nodes from the different clusters
(Line~\ref{alg:da:newdistance}).
If the new cluster is still $\DEPTHTHRES$-depth and
$\DISTTHRES$-distance consistent
(Lines~\ref{alg:da:depthdistancecheck}), we merge the current
candidate data area into $\Last$ and continue
(Line~\ref{alg:da:domerge}).

Otherwise, the cluster $\Last$ cannot be grown further. Then, if
$\Last$ contains at least 2 nodes (Line~\ref{alg:da:sizecheck}), we
compute the representative $d$ of $\Last$ as the least common ancestor
$\lca(\Nodes_{\Last})$ of the contained pivot nodes $\Nodes_{\Last}$
(Line~\ref{alg:da:lca}).
If this representative $d$ is not already bound to another (earlier
occurring) support set of at least of the same size
(Line~\ref{alg:da:nolargerdacheck}), we assign $\Nodes_{\Last}$ as new
support to $\PIVOTS(d)$ and mark $d$ as dataarea by adding
$\TYPING(d,\dataareatype)$ (Line~\ref{alg:da:doadd}).
At last, we start a to build a data area with the current one
$(\Nodes,\Depth,\Dist)$.
The algorithm always enters this else branch during its last iteration
to ensure that the very last data area's pivot nodes are properly
considered as a possible support set.

\begin{compacttheo}
  The set of data areas for a DOM $P$ of size $n$ under schema
  $\Sigma$ and pivot attribute type $\pi$ is computed in $O(n^2)$.
\end{compacttheo}

\begin{compactproof}  
  Lines~\ref{alg:da:initpivots}--\ref{alg:da:initlastda} iterate twice
  over the DOM and are therefore in $O(n)$.
  Lines~\ref{alg:da:loopcandidates}--\ref{alg:da:lastdareset} are in
  $O(n^2)$, as the loop is dominated by the computation of the
  distance intervals.
  For the distance intervals, we extend the interval by the maximum
  and minimum path length between nodes from $\Nodes_{\Last}$ and
  $\Nodes$ and thus compare any pair of nodes at most once (when
  merging it to the previous cluster).\qed
\end{compactproof}


\begin{figure}[tbp]
  \centering
  \includegraphics[width=\linewidth]{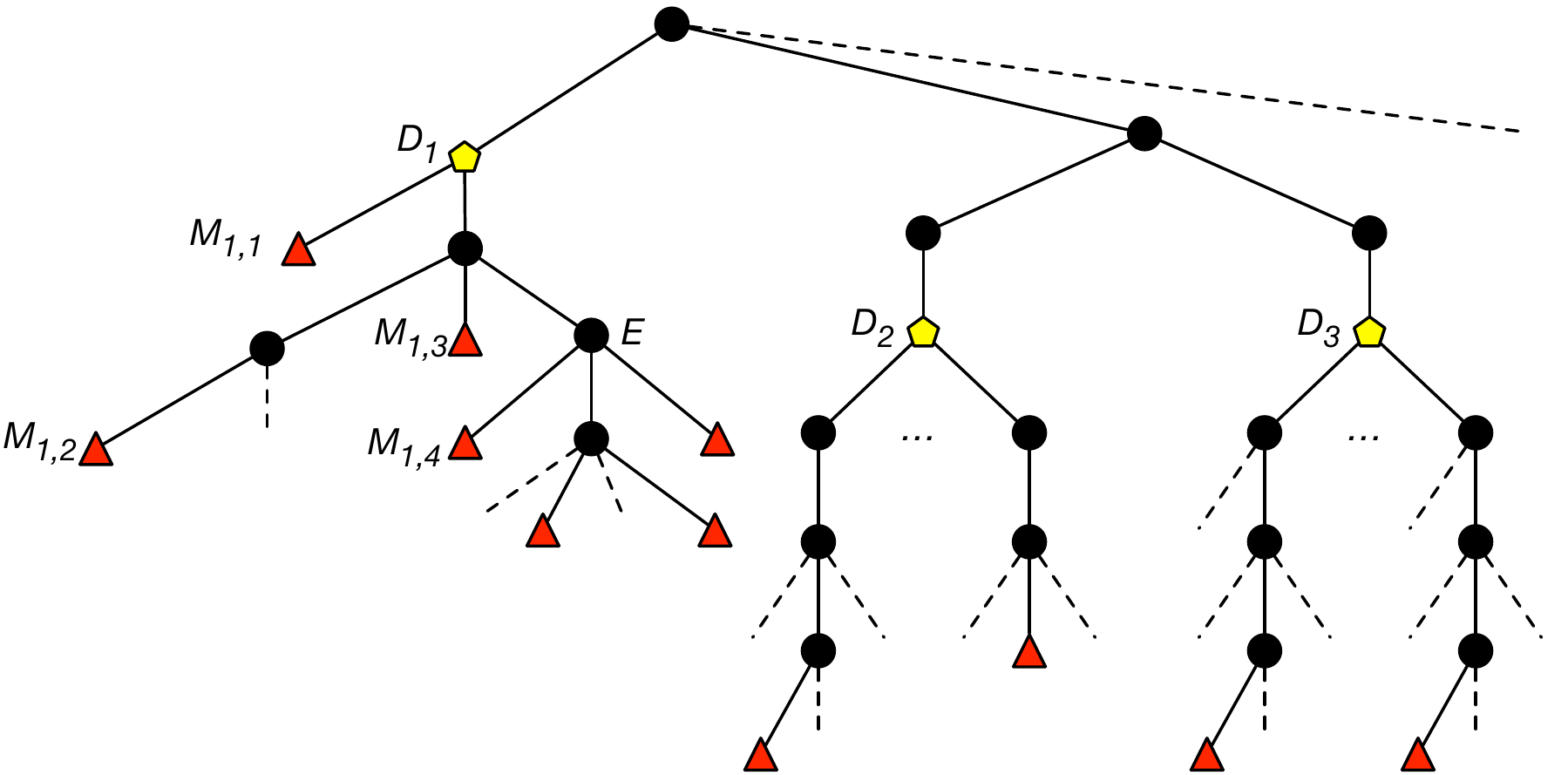}
  \caption{Data area identification}
  \label{fig:example-data-area}
\end{figure}

To illustrate Algorithm~\ref{algo:data-area}, consider
Figure~\ref{fig:example-data-area} with $\DISTTHRES = \DEPTHTHRES =
3$.
Yellow diamonds represent the data areas $D_1, D_2$, and $D_3$, and
red triangles pivot nodes.
With this large thresholds the algorithm creates one cluster at $D_1$
with $M_{1,1}$ to $M_{1,4}$ as support, despite the homogeneity of the
subtree rooted at $E$ and the ``loss'' of the three rightmost pivot
nodes in $E$.
In Section~\ref{sec:evaluation}, we show that the best results are
obtained with smaller thresholds, viz.\@ $\DISTTHRES =2$ and
$\DEPTHTHRES = 1$, which indeed would split $D_1$ in this case.
Also note, that $D_2$ and $D_3$ are not distance consistent and thus
cannot be merged.
Small variations in depth and distance, however, such as in $D_2$ do
not affect the data area identification.

\begin{figure*}[tbp]
  \centering \includegraphics[width=1\linewidth]{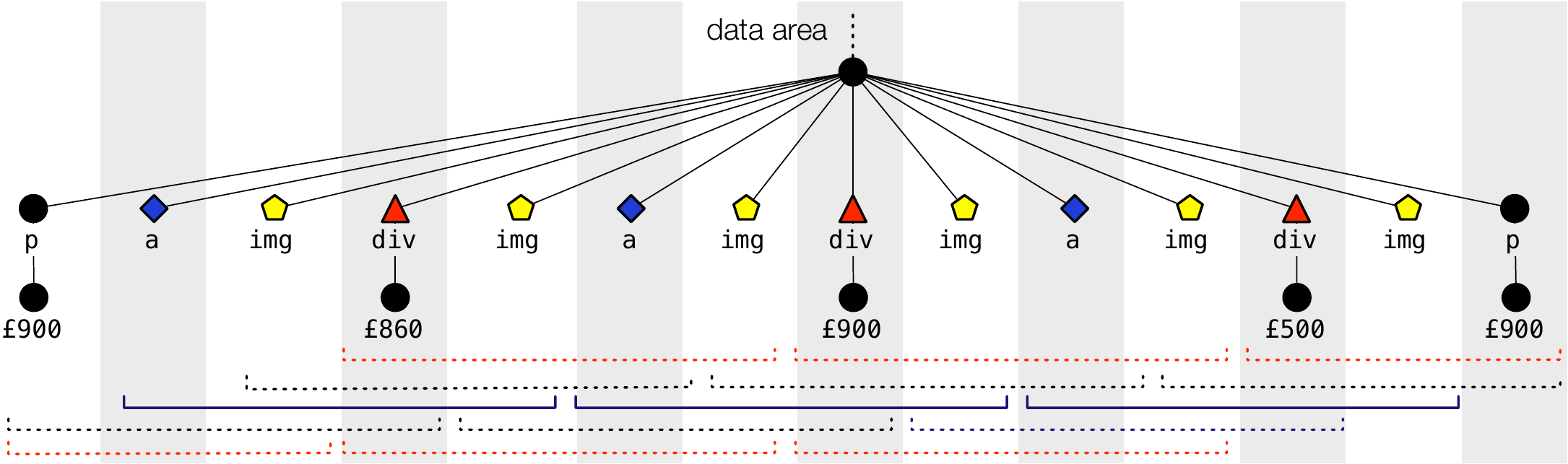}
  \caption{ Record Segmentation}
  \label{fig:complex-segmentation}
\end{figure*}

\subsection{Record Segmentation}
\label{subsec:segmentation}

During the data area identification, \AMBER identifies data areas of a
page, marks their roots $d$ with $\TYPING(d,\dataareatype)$, and
provides the pivot nodes $\PIVOTS(d)$ supporting the data area, with
its pivot nodes occurring roughly at the same depth and mutual
distance.
As in data area identification, \AMBER approximates the occurrence of
relevant data and structural record similarity through instances of
regular attribute types \REGATTS\ {\bfseries(R4)} to construct a set of
candidate segmentations.
Hence, only the records in these candidate segmentations must be
checked for mutual structural similarity {\bfseries(R3)}, allowing
\AMBER to scale to large and complex pages at ease.

\begin{compactdef}
  A \textbf{record} is a set $r$ of continuous children of a data area
  $d$ {\bfseries(R1)}, such that $r$ contains at least one pivot node
  from $\PIVOTS(d)$ {\bfseries(R4)}.
  A \textbf{record segmentation} of $d$ is a set of non-overlapping
  records $\RECORDS$ of uniform size {\bfseries(R2)}.
  The quality of a segmentation $\RECORDS$ improves with increasing
  size {\bfseries(D1)} and decreasing irregularity {\bfseries(R3)}.
\end{compactdef}

Given a data area root $d$ and its pivot nodes $\PIVOTS(d)$, this
leads to a dual objective optimization problem, striving for a maximal
area of minimal irregularity. 
We concretize this problem with the notion of \emph{leading nodes:}
Given a pivot node $n\in\PIVOTS(d)$, we call the child $l$ of $d$,
containing $n$ as a descendant, the leading node $l$ of $n$.
Accordingly, we define $\LEADINGS(d)$ as the set of leading nodes of a
data area rooted at $d$.
To measure the number of siblings of $l$ potentially forming a record,
we compute the \emph{leading space} $\lspace(l,L)$ after a leading
node $l\in L$ as the sibling distance $l \SDIST l'$, where $l'\in L$
is the next leading node in document order. 
The two objectives for finding an optimal record segmentation
\RECORDS are then as follows:

\begin{compactenum}[\bfseries(1)]
\item Maximize the subset $\RECORDS'\subseteq\RECORDS$ of records that
  are \emph{evenly segmented} {\bfseries(D1)}.
  A subset $\RECORDS'=\{r_1,\dots r_k\}$ is evenly segmented if each
  record $r_i\in\RECORDS'$ contains exactly one pivot node
  $n_i\in\PIVOTS(d)$ {\bfseries(R4)}, and
  all leading nodes $l_i$ corresponding to a pivot node $n_i$ have the
  same leading space $\lspace(l_i,\LEADINGS(d))$ {\bfseries(R1-3)}.
\item Minimize the irregularity of the record segmentation
  {\bfseries(R3)}.
  The \emph{irregularity} of a record segmentation $\RECORDS$
  equals the summed relative tree edit distances between all pairs of
  nodes in different records in $\RECORDS$, i.e.,
  $\irregularity(\RECORDS) = \sum_{n \in r, n' \in r' \text{with }
    r \neq r' \in \RECORDS}\editDistance(n,n')$, where
  $\editDistance(n,n')$ is the standard tree edit distance normalized
  by the size of the subtrees rooted at $n$ and $n'$ (their
  ``maximum'' edit distance).
\end{compactenum}

\AMBER approximates such a record segmentation with
Algorithm~\ref{alg:partitioning}.
It takes as input a DOM $P$, a data area root $d\in P$, and accesses
the corresponding support sets via $\PIVOTS(d)$, as constructed by the
data area identification algorithm of the preceding section.
The segmentation is computed in two steps, first searching a basic
record segmentation that contains a large sequence of evenly segmented
pivot nodes, and second, shifting the segmentation boundaries back and
forth to minimize the irregularity.
In a preprocessing step all children of the data area without text or
attributes (``empty'' nodes) are collapsed and excluded from the
further discussion, assuming that these act as separator nodes, such
as \htmltag{br} nodes.

So, the algorithm initially determines the sequence $\mathcal{L}$ of
leading nodes underlying the segmentation (Line~\ref{alg:seg:leading}).
Based on these leading nodes, the algorithm estimates the distance
$\textsf{Len}$ between leading nodes (Line~\ref{alg:seg:len}) that
yields the largest evenly segmented sequence:
We take for $\textsf{Len}$ the shortest leading space $\lspace(l)$
among those leading spaces occurring most often in $\mathcal{L}$.
Then we deal with noise prefixes in removing those leading nodes $l_k$
from the beginning of $\mathcal{L}$ which have $\lspace(l_k)$ smaller
than $\textsf{Len}$
(Line~\ref{alg:seg:headleadingcleanloop}-\ref{alg:seg:headleadingstep}).
After dealing with the prefixes, we drop all leading nodes from
$\mathcal{L}$ whose sibling distance to the previous leading node is
less than $\textsf{Len}$
(Lines~\ref{alg:seg:leadingcleanloop}-\ref{alg:seg:leadingstep}).
This loop ensures that each remaining leading node has a leading space
of at least $\textsf{Len}$ and takes care of noise suffixes.

With the leading nodes $\mathcal{L}$ as a frame for segmenting the
records, the algorithm generates all segmentations with record size
$\textsf{Len}$ such that each record contains at least one leading
node from $\mathcal{L}$.
To that end, the algorithm computes all possible sets $\StartCands$ of
record start points for these records by shifting the original leading
nodes $\mathcal{L}$ to the left (Line~\ref{alg:seg:startcandidates}).
The optimal segmentation $\OptimalPartition$ is set to the empty set,
assuming that the empty set has high irregularity
(Line~\ref{alg:seg:loopinit}).
We then iterate over all such start point sets $S$
(Line~\ref{alg:seg:startcandloop}) and compute the actual
segmentations $\Partition$ as the records of $\textsf{Len}$ length,
each starting from one starting point in $S$
(Line~\ref{alg:seg:segmentation}).
By construction, these are records, as they are continuous siblings
and contain at least one leading node (and hence at least one pivot
node).
The whole $\textsf{Segmentation}$ is a record segmentation as its
records are non-overlapping (because of
Line~\ref{alg:seg:leadingcleanloop}-\ref{alg:seg:leadingstep}) and of
uniform size $\textsf{Len}$
(Line~\ref{alg:seg:segmentationsizecheck}).
From all constructed segmentations, we choose the one with the lowest
irregularity
(Lines~\ref{alg:seg:segmentationoptimalitycheck}-\ref{alg:seg:segmentationtaken}).
At last, we iterate through all records $r$ in the optimal
segmentation $\OptimalPartition$ (Line~\ref{alg:seg:labelloop}), and
mark the first node $n\in r$ as record start with
$\TYPING(n,\recordstarttype)$ (Line~\ref{alg:seg:labelfirstassign})
and all remaining nodes $n\in r$ as record tail with
$\TYPING(n,\recordtailtype)$ (Line~\ref{alg:seg:labeltailassignloop}-\ref{alg:seg:labeltailassign}).

\begin{algorithm}[tbp]\small
\SetKw{new}{new}
\SetKw{LET}{let}
\SetKw{DELETE}{delete}
\SetKw{DELETESKIP}{delete-skip}
\SetKwFunction{add}{add}
\SetKwFunction{recordLength}{recordLength}
\SetKwFunction{IdentifyLeadings}{IdentifyLeadings}
\SetKwData{Children}{Children}
\SetKwData{Root}{root}
\SetKwData{RecSize}{Len}
\SetKwData{OptimalSim}{MinIrregularity}
\SetKwData{Segmentation}{Segmentation}
\SetKwData{ShallowSimilarity}{ShallowSimilarity}
\SetKwData{SubTreeSimilarity}{SubTreeSimilarity}

\Input{$P$ -- DOM to be analyzed}
\Input{$\TYPING$ -- extraction typing on $P$}
\Input{$\PIVOTS$ -- data areas support}
\Modifies{$\TYPING$ -- adds record segmentations}

\ForEach{$d\in P : \TYPING(d,\dataareatype)$}
{
$\mathcal{L} \gets \LEADINGS(d)$\;\label{alg:seg:leading}
$\RecSize \gets \min\{\lspace(l,\mathcal{L})$: $l\in\mathcal{L}$ with maximal
$|\{l'\in\mathcal{L}: \lspace(l,\mathcal{L}) = \lspace(l',\mathcal{L})\}|\}$\;\label{alg:seg:len}

\While{$l_k\in\mathcal{L}$ in document order, $\lspace(l_k,\mathcal{L}) < \RecSize$}{\label{alg:seg:headleadingcleanloop}
  \DELETE $l_{k}$ from $\mathcal{L}$\;\label{alg:seg:headleadingstep}
}
\For{$l_k\in\mathcal{L}$ in document order}{\label{alg:seg:leadingcleanloop}
  \lIf{$\lspace(l_k,\mathcal{L}) < \RecSize$}{
     \DELETESKIP $l_{k+1}$ from $\mathcal{L}$\;\label{alg:seg:leadingstep}
   }
}
$\StartCands \gets \{\{n: \exists l \in
  \mathcal{L}: n \SDIST l = i\}: 0\leq i < \RecSize\}$\;\label{alg:seg:startcandidates}
 $\OptimalPartition \gets \emptyset$\;\label{alg:seg:loopinit}
\ForEach{$S \in \StartCands$}{\label{alg:seg:startcandloop}
  $\Partition=\{ \{n: s \SDIST n \leq \RecSize\}\; : \; s\in S\}$\;\label{alg:seg:segmentation}
  \If{$\forall r \in \Partition: |r| = \RecSize$}{\label{alg:seg:segmentationsizecheck}
    \If{$\irregularity(\Partition) < \irregularity(\OptimalPartition)$}{\label{alg:seg:segmentationoptimalitycheck}
      $\OptimalPartition \gets \Partition$\; \label{alg:seg:segmentationtaken}
    }
  }
}
\ForEach{$r\in\OptimalPartition$}{\label{alg:seg:labelloop}
   \ForEach{Node $n_i\in r$ in document order}{\label{alg:seg:labeltailassignloop}
     \Add $\TYPING(n_i,\recordtailtype)$\;\label{alg:seg:labeltailassign}}
   \Add $\TYPING(n_1,\recordstarttype)$\;\label{alg:seg:labelfirstassign}
}}
\caption{\recordalgo$(P, \TYPING, \PIVOTS$)}
\label{alg:partitioning}
\end{algorithm}

\begin{compacttheo}
  Algorithm~\ref{alg:partitioning} runs in $O(b \cdot n^3)$ on a data
  area $d$ with $b$ as degree of $d$ and $n$ as size of the subtree
  below $d$.
\end{compacttheo}

\begin{compactproof}
  Lines~\ref{alg:seg:leading}-\ref{alg:seg:startcandidates} are in
  $O(b^2)$.
  Line~\ref{alg:seg:startcandidates} generates in \StartCands at most
  $b$ segmentations (as $\textsf{Len} \leq b$) of at most $b$ size.
  The loop in
  Lines~\ref{alg:seg:startcandloop}-\ref{alg:seg:segmentationtaken} is
  executed once for each segmentation $S\in\StartCands$ and is
  dominated by the computation of $\irregularity()$ which is bounded
  by $O(n^3)$ using a standard tree edit distance algorithm.
  Since $b \leq n$, the overall bound is $O(b^2+b \cdot n^3=b\cdot n^3)$.\qed
\end{compactproof}


In the example of Figure~\ref{fig:complex-segmentation}, \AMBER
generates five segmentations with $\textsf{Len}=4$, because of the
three (red) \htmltag{div} nodes, occurring at distance 4.
Note, how the first and last leading nodes (\htmltag{p} elements) are
eliminated (in Lines~\ref{alg:seg:headleadingcleanloop}-\ref{alg:seg:leadingstep}) as
they are too close to other leading nodes.
Of the five segmentations (shown at the bottom of
Figure~\ref{fig:complex-segmentation}), the first and the last are
discarded in Line~\ref{alg:seg:segmentationsizecheck}, as they contain
records of a length other than $4$.
The middle three segmentations are proper record segmentations, and
the middle one (solid line) is selected by \AMBER, because it has the
lowest irregularity among those three.

\subsection{Attribute Alignment}
\label{subsec:attr-reconciliation}

After segmenting the data area into records, \AMBER aligns the contained
attributes to complete the extraction instance.
We limit our discussion to single valued attributes, i.e., attribute
types which occur at most once in each record.
In contrast to other data extraction approaches, \AMBER does not need
to refine records during attribute alignment, since the repeating
structure of attributes is already established in the extraction
typing. 
It remains to align all attributes with sufficient cross-record
support, thereby inferring missing attributes, eliminating noise ones,
and breaking ties where an attribute occurs more than once in a single
record.

When aligning attributes, \AMBER must compare the position of
attribute occurrences in different records to detect repeated
structures {\bfseries(R3)} and to select those attribute instances
which occur at similar relative positions within the records
{\bfseries(R6)}.
To encode the position of an attribute relative to a record, we use
the path from the record node to the attribute:

\begin{compactdef}
  For DOM nodes $r$ and $n$ with $\DOMREL{descendant}(r,n)$, we define
  the \textbf{characteristic tag path} $\TAGPATH_{r}(n)$ as the
  sequence of HTML tags occurring on the path from $r$ to $n$,
  including those of $r$ and $n$ itself, taking only
  \DOMREL{first-child} and \DOMREL{next-sibl} steps while skipping all
  text nodes. With the exception of $r$'s tag, all HTML tags are
  annotated by the step type.
\end{compactdef}

\begin{figure}[tbp]
  \centering
  \includegraphics[width=\linewidth]{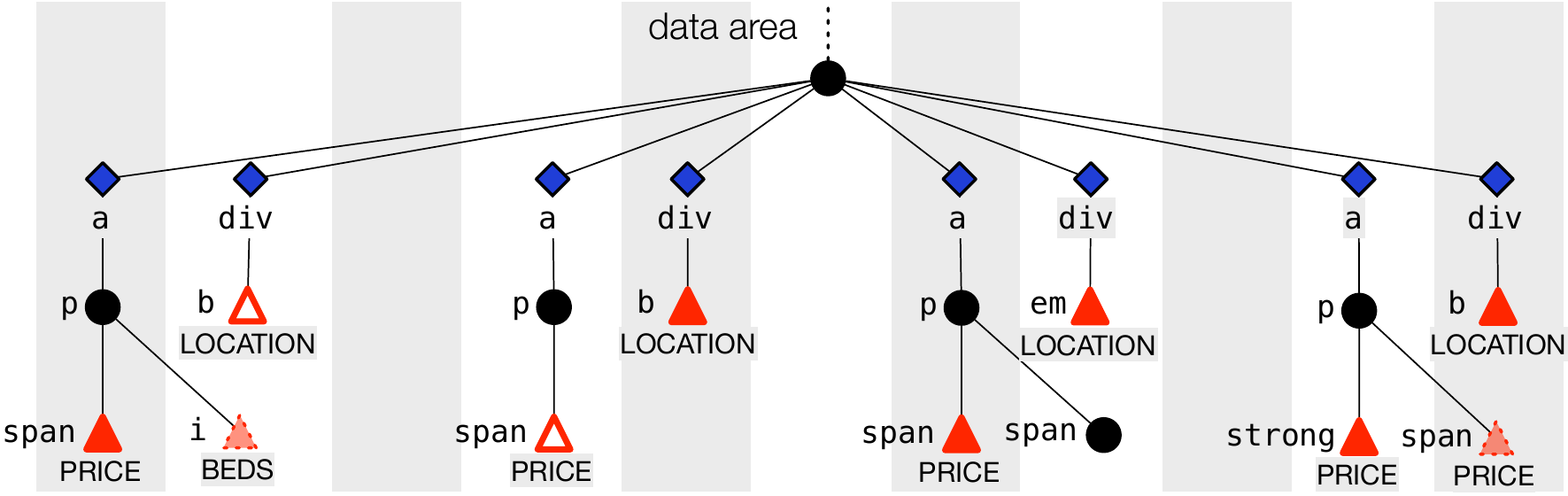}
  \caption{Attribute alignment}
  \label{fig:example-attributes}
\end{figure}

For example, in Figure~\ref{fig:example-attributes}, the
characteristic tag path from the leftmost \htmltag{a} and to its
\htmltag{i} descendant node is
\texttt{a/first-child::p/first-child::span/next-sibl::i}.
Based on characteristic tag paths, \AMBER quantifies the assumption
that a node $n$ is an attribute of type $\rho\in\Sigma$ within record
$r$ with \emph{support} $\SUPPORT_{\TYPING}(r,n,\rho)$.

\begin{compactdef}
  Let $\TYPING$ be an extraction typing on DOM $P$ with nodes
  $d,r,n\in P$ where $n$ belongs to record $r$, and $r$ belongs to the
  data area rooted at $d$.
  Then the \textbf{support} $\SUPPORT_{\TYPING}(r,n, \rho)$ for $n$ as
  attribute instance of type $\rho\in\SCHEMA$ is defined as the
  fraction of records $r'$ in $d$ that contain a node $n'$ with
  $\TAGPATH_r(n)=$ $\TAGPATH_{r'}(n')$ and $\TYPING(n',\rho)$ for
  arbitrary $v$.
\end{compactdef}

Consider a data area with 10 records, containing 1
$\type{price}$-annotated node $n_1$ with tag path
\DOMREL{div/\dots/next-sibl::span} within record $r_1$, and 3
$\type{price}$-annotated nodes $n_2\dots n_4$ with tag path
\DOMREL{div/\dots/first-child::p} within records $r_2\dots r_4$, resp.
Then, $\SUPPORT_\TYPING(r_1,n_1,\type{price}) = 0.1$ and
$\SUPPORT_\TYPING(r_i,n_i,\type{price}) = 0.3$ for $2\le i\le 4$.

With the notion of support at hand, we define our criterion for an
acceptable extraction typing \TYPING~-- which we use to transform
incomplete and noise annotations into consistent attributes: 
We turn annotations into attributes if the support is strong enough,
and with even stronger support, we also infer attributes without
underlying annotation.
\begin{compactdef}
  \label{sec:well-supported}
  An extraction typing \TYPING over schema
  $\SCHEMA=\REGATTS\cup\OPTATTS$ and DOM $P$ is
  \textbf{well-supported}, if for all nodes $n$ with
  $\TYPING(n,\rho)$, one of the following two conditions is satisfied~--
  setting $X=R$ for $\rho\in\REGATTS$ and $X=O$ for $\rho\in\OPTATTS$:
  \begin{inparaenum}[\bfseries(1)]
  \item $\SUPPORT_\TYPING(r,n,\rho) > \InventThres_X$, or
  \item $\SUPPORT_\TYPING(r,n,\rho) > \DeleteThres_X$ and
    $\ANNOT(\rho,n,v)$.
  \end{inparaenum}
\end{compactdef}
This definition introduces two pairs thresholds,
$\RegularInventThres$, $\RegularDeleteThres$ and
$\OptionalInventThres$, $\OptionalDeleteThres$, respectively, for
dealing with regular and optional attribute types.
In both cases, we require $\InventThres_X>\DeleteThres_X$, as
inferring an attribute without an annotation requires more support
than keeping a given one.
We also assume that $\RegularInventThres\geq\OptionalInventThres$,
i.e., that optional attributes are easier inferred, since optional
attributes tend to come with more variations (creating false
negatives) {\bfseries(A3}).
Symmetrically, we assume
$\OptionalDeleteThres\geq\RegularDeleteThres$, i.e., that optional
attributes are easier dropped, optional attributes that are not cover
by the template {\bfseries(R5)} might occur in free-text descriptions
(creating false positives).
Taken together, we obtain $\RegularInventThres \geq
\OptionalInventThres \geq \OptionalDeleteThres \geq
\RegularDeleteThres$.
See Section~\ref{sec:evaluation} for details on how we set these four
thresholds.

We also apply a simple pruning technique prioritizing early
occurrences of attributes {\bfseries(A1)}, as many records start with
some semi-structured attributes, followed by a free-text description.
Thus earlier occurrences are more likely to be structured attributes
rather than occurrences in product descriptions.
%
%
As shown in Section~\ref{sec:evaluation}, this simple heuristic
suffices for high-accuracy attribute alignment. For clarity and space
reasons, we therefore do not discuss more sophisticated attribute
alignment techniques.

\begin{algorithm}[tbp]\small
\SetKw{LET}{let}
\SetKw{DELETE}{delete}
\SetKw{ADD}{add}
\SetKwData{Mandatories}{Mandats}
\SetKwFunction{FieldScopeLabelling}{FieldScopeLabelling}
\SetKwFunction{id}{id}
\SetKwFunction{computeSupport}{computeSupport}
\SetKw{Break}{break}
\SetKw{Continue}{continue}
\SetKwFunction{AddAttribute}{addAttribute}
\SetKwFunction{RemoveAttribute}{removeAttribute}
\SetKwFunction{RemoveAttributes}{removeAttributes}
\SetKwFunction{ReplaceAttribute}{replaceAttribute}
\SetKwFunction{RemoveRecord}{removeRecord}
\SetKwFunction{isMandatory}{isMandatory}
\SetKwFunction{isForbidden}{isForbidden}
\SetKwFunction{LookupAttributesPerType}{LookupAttributesPerType}
\SetKwFunction{rankBySupport}{RankAttributesBySupport}
\SetKwFunction{Missing}{MissingAttribute}

\Input{$P$ -- DOM to be analyzed}
\Input{$\TYPING$ -- extraction typing on $P$}
\Input{$\SCHEMA=\REGATTS\cup\OPTATTS$\ -- schema of the searched results}
\Modifies{$\TYPING$ -- adds attributes}

\ForEach{$\rho$ in $\Sigma$}{\label{alg:align:attributeloop}
  \Select $X$ \With $\rho\in\Sigma_X:$   $\InventThres \gets \InventThres_X$;\label{alg:align:thresholds}
    $\DeleteThres \gets \DeleteThres_X$\;
  \ForEach{$n,r\in P$ \With  $\partOf_\TYPING(n,r)$}{\label{alg:align:nodeloop}
    \lIf{$\SUPPORT_\TYPING(r,n,\rho)>\InventThres$  \label{alg:align:ifinfer}
      \Or $\big(\ANNOT(\rho,n,v)$ \LogicalAnd $\SUPPORT_E(r,n,\rho) > \DeleteThres\big)$}{
      \Add $\TYPING(n,\rho)$\;
    }
  }
  \ForEach{$n$ \With $\rho\in\TYPING(n)$}{\label{alg:align:typednodeloop}
    \If{$\exists n':$ $\TYPING(n',\rho)$
      \LogicalAnd $\DOMREL{following}(n',n)$}{\label{alg:align:disambiguateif}
      \Remove $\TYPING(n,\rho)$\;\label{alg:align:disambiguate}
    }
  }
}
\caption{\attributealgo$(P, \TYPING, \Sigma)$}
\label{alg:reconciliation}
\end{algorithm}

Algorithm~\ref{alg:reconciliation} shows the full attribute alignment
algorithm and presents a direct implementation of the
well-supportedness requirement.
The algorithm iterates over all attributes in the schema
$\rho\in\SCHEMA=\REGATTS\cup\OPTATTS$
(Line~\ref{alg:align:attributeloop}) and selects the thresholds
$\InventThres$ and $\DeleteThres$ depending on whether $\rho$ is
regular or optional (Line~\ref{alg:align:thresholds}).
Next, we iterate over all nodes $n$ which are part of a record $r$
(Line~\ref{alg:align:nodeloop}). 
We assign the attribute type $\rho$ to $n$, if the support
$\SUPPORT_\TYPING(r,n,\rho)$ for $n$ having type $\rho$ is reaching
either the inference threshold $\InventThres$ or the keep threshold
$\DeleteThres$, requiring additionally an annotation
$\ANNOT(\rho,n,v)$ in the latter case (Line~\ref{alg:align:ifinfer}).
After finding all nodes $n$ with enough support to be typed with
$\rho$, we remove all such type assignments except for the first one
(Lines~\ref{alg:align:typednodeloop}-\ref{alg:align:disambiguate}).

\begin{compacttheo}
  \AMBER's attribute alignment (Algorithm~\ref{alg:reconciliation})
  computes a well-supported extraction instance for a page with DOM
  $P$ in $O(|\Sigma| \cdot |P|)$.
\end{compacttheo}

In Figure~\ref{fig:example-attributes} we illustrate attribute
alignment in \AMBER for $\InventThres=40\%$ for both regular and
optional attribute types and $\RegularDeleteThres=0\%$,
$\OptionalDeleteThres=30\%$ (\TYPE{price} and \TYPE{location} regular,
\TYPE{beds} optional): The data area has four records each spanning
two of the children of the data area (shown as blue diamonds).
Red triangles represent attributes with the attribute type written
below.
Other labels are HTML tags.
A filled triangle is an attribute directly derived from an annotation,
an empty triangle one inferred by the algorithm in Line $6$.
In this example, the second record has no \type{price}
annotation.
However, there is a \htmltag{span} with tag path
\texttt{a/first-child::p/first-child::span} and there are two other
records (the first and third) with a \htmltag{span} with the same tag
path from their record.
Therefore that \htmltag{span} has support $>\InventThres=40\%$ for
\TYPE{price} and is added as a \TYPE{price} attribute to the second
record.
Similarly, for the \htmltag{b} element in record $1$ we infer type
\TYPE{location} from the support in record $2$ and $4$.
Record $3$ has a \type{location} annotation, but in an
\htmltag{em}.
This has only $25\%$ support, but since \TYPE{location} is regular
that suffices.
This contrasts to the \htmltag{i} in record $1$ which is annotated as
\TYPE{beds} and is not accepted as an attribute since optional
attributes need at least $\OptionalDeleteThres=30\%$ support.
In record $4$ the second \type{price} annotation is ignored since it
is the second in document order (Lines 7--8).

\subsection{Running Example}
\label{subsec:running-example2}

Recall Figure~\ref{fig:rightmove-example} in
Section~\ref{sec:running-example}, showing the web page of
\texttt{rightmove.co.uk}, an UK real estate aggregator, which we use
as running example:
It shows a typical result page with one data area with featured
properties {\bfseries(1)}, a second area with regular search results
{\bfseries(2)}, and a menu offering some filtering options
{\bfseries(3)}.

\begin{figure}[tbp]
  \centering
  \includegraphics[width=\linewidth]{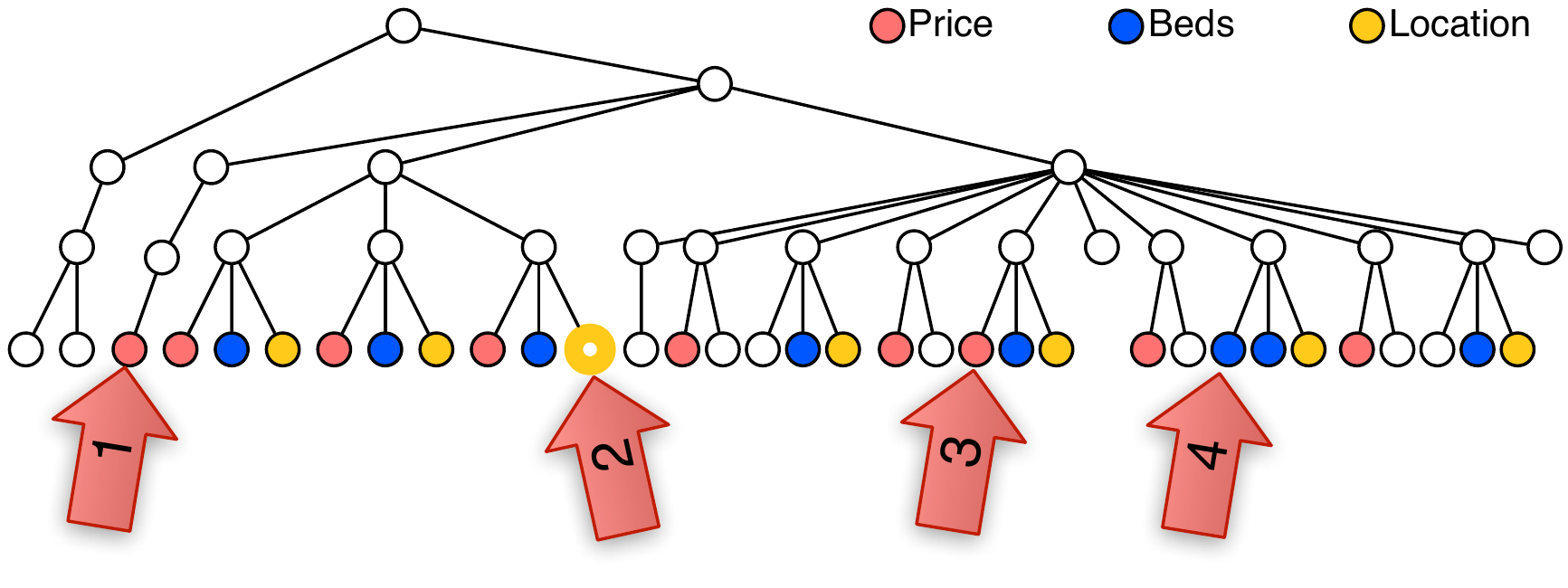}
  \caption{Simplified DOM for \texttt{rightmove.co.uk} }
 \label{fig:rightmove-example-dom}
\end{figure}

For this web page, Figure~\ref{fig:rightmove-example-dom} shows a
simplified DOM along with the raw annotations for the attribute types
\type{price}, \type{bedRoomNumber}, and \type{location}, as provided
by our annotation engine (for simplicity, we do not consider the
\type{bathRoomNumber} shown on the original web page).
Aside the very left nodes in Figure~\ref{fig:rightmove-example-dom},
belonging to the filter menu, the DOM consists of a single large
subtree with annotated data. 
The numbered red arrows mark noise or missing annotations~-- to be
fixed by \AMBER:
\begin{inparaenum}[\bfseries(1)]
\item This node contains indeed a price, but outside any record: It is
  the average rent over the found results, occurring at the very top
  of Figure~\ref{fig:rightmove-example}.
\item The \type{location} annotation in the third record is missing.
\item The second price in this record is shown crossed out, and is
  therefore noise to be ignored.
\item This bedroom number refers to a flat to sublet within a larger
  property and is therefore noise.
\end{inparaenum}

\begin{figure}[tbp]
  \centering
  \includegraphics[width=\linewidth]{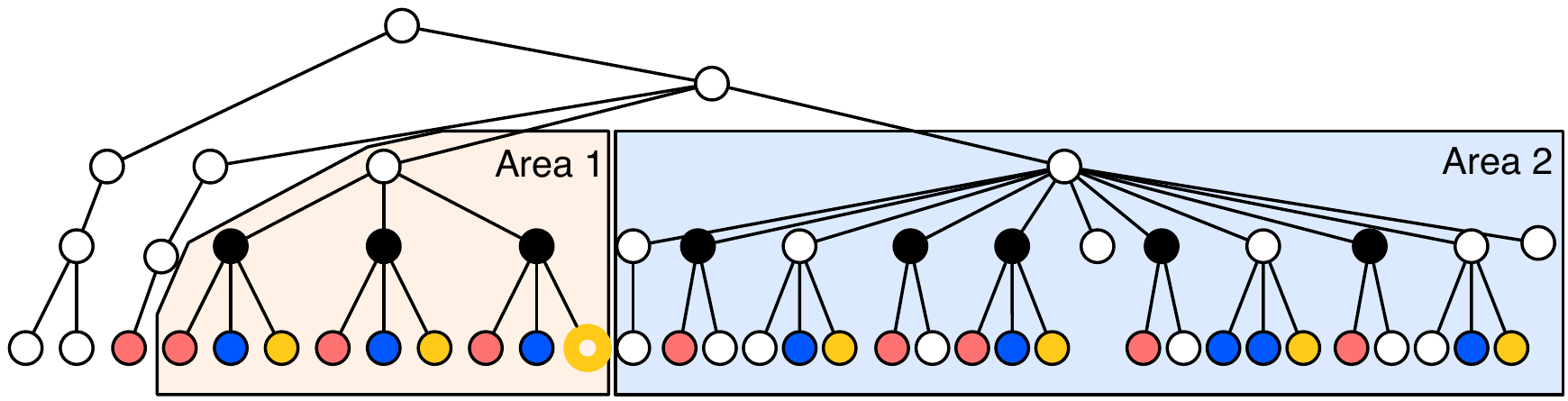}
  \caption{Data area identification on \texttt{rightmove.co.uk} }
 \label{fig:rightmove-example-da}
\end{figure}

\paragraph*{Data Area Identification.}\ 
\label{sec:data-area-ident}
For identifying the data areas, shown in
Figure~\ref{fig:rightmove-example-da}, Algorithm~\ref{algo:data-area}
searches for instances of the pivot attribute type~-- \type{price} in
this case.
\AMBER clusters all pivot nodes which are depth and distance
consistent for $\DEPTHTHRES=\DISTTHRES=1$ into one data area, obtaining
the shown Areas 1 and 2.
The \type{price} instance to the very left (issue {\bfseries(1)} named
above) does not become part of a cluster, as it its distance to all
other occurrences is 6, whereas the occurrence inside the two clusters
have mutual distance 4, with $4+\DISTTHRES<6$.
For same reason, the two clusters are not merged, as the distance
between one node from Area 1 and one from Area 2 is also 6.
The data area is then identified by the least common ancestor of the
supporting pivot nodes, called the data area root.

\begin{figure}[tbp]
  \centering
  \includegraphics[width=\linewidth]{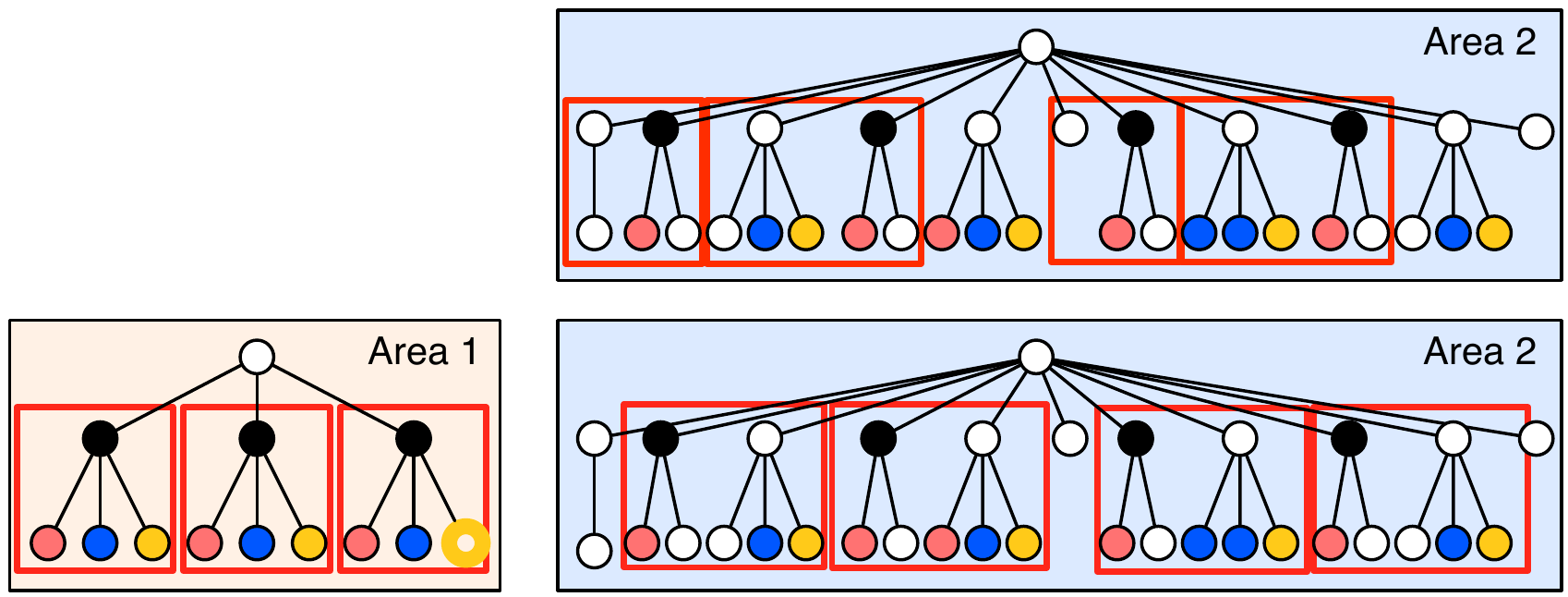}
  \caption{Record Segmentation on \texttt{rightmove.co.uk} }
 \label{fig:rightmove-example-rs}
\end{figure}

\paragraph*{Record Segmentation.}\ 
\label{sec:record-segmentation}
The record segmentation in Algorithm~\ref{alg:partitioning} processes
each data areas in isolation:
For a given area, it first determines the leading nodes corresponding
to the pivot nodes, shown as solid black nodes in
Figure~\ref{fig:rightmove-example-da}. 
The leading node of a pivot node is the child of the data area root
which is on the path from the area root to the pivot node.
In case of Area 1 to the left, all children of the area root are
leading nodes, and hence, each subtree rooted at a leading nodes
becomes a record in its own right, producing the segmentation shown to
the left of Figure~\ref{fig:rightmove-example-rs}.
The situation within Area 2 is more complicated: \AMBER first
determines the record length to be 2 sibling children of the area
root, since in most cases, the leading nodes occur in a distance of 2,
as shown in Figure~\ref{fig:rightmove-example-da}.
Having fixed the record length to 2, \AMBER drops the leading nodes
which follow another leading node too closely, eliminating the leading
node corresponding to the noisy \type{price} in the second record
(issue {\bfseries(3)} from above).
Once the record length and the resulting leading nodes are fixed,
Algorithm~\ref{alg:partitioning} shifts the records boundaries to find
the right segmentation, yielding two alternatives, shown on the right
of Figure~\ref{fig:rightmove-example-rs}. 
In the upper variant, only the second and fourth record are similar,
the first and third record deviate significantly, causing a lot of
irregularity.
Hence, the lower variant is selected, as its four records have a
similar structure.

\begin{figure}[tbp]
  \centering
  \includegraphics[width=\linewidth]{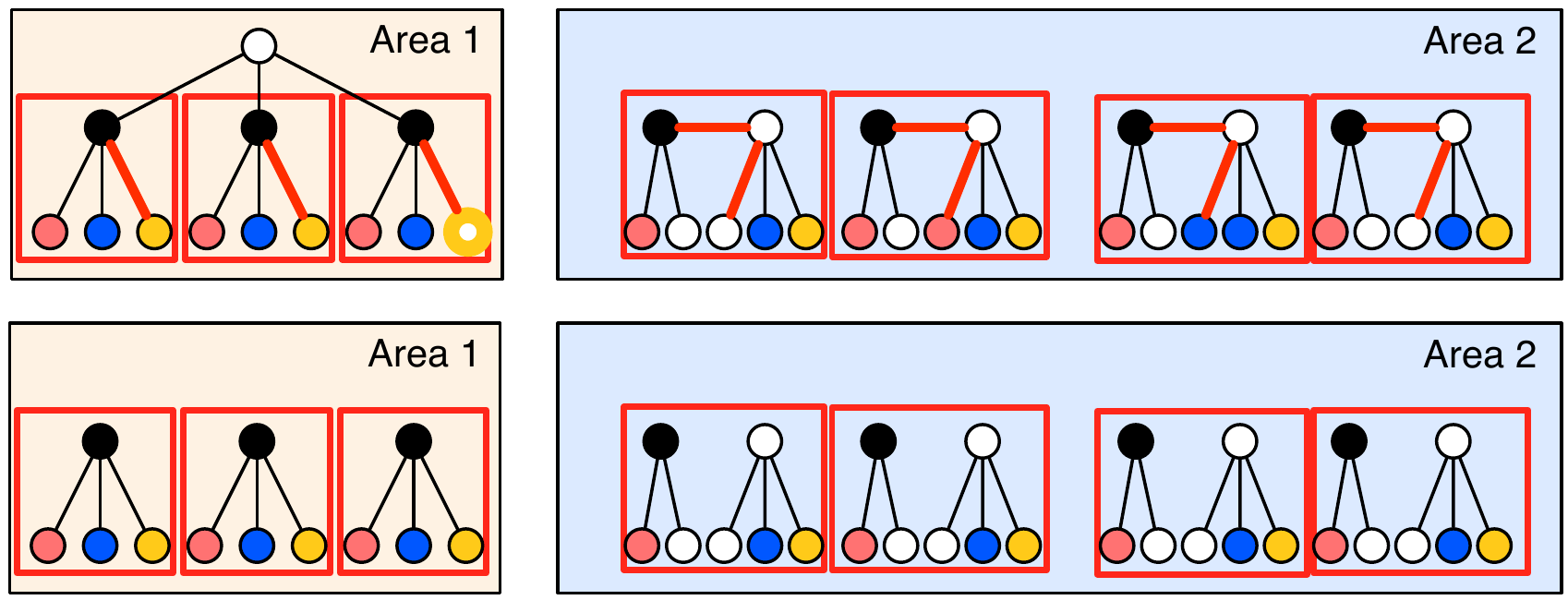}
  \caption{Attribute Alignment on \texttt{rightmove.co.uk} }
 \label{fig:rightmove-example-aa}
\end{figure}

\paragraph*{Attribute Alignment.}\ 
\label{sec:attribute-alignment}
Algorithm~\ref{alg:reconciliation} fixes the attributes of the
records, leading to the record structure shown in lower half of Figure
\ref{fig:rightmove-example-aa}. 
This infers the missing \type{location} and cleans the noisy
\type{price} (issues {\bfseries (2)} and {\bfseries(4)} from above).
One the upper left of Figure~\ref{fig:rightmove-example-aa}, we show
the characteristic tag path for \type{location} is computed, resulting
in a support of $2/3$, as we have 2 \type{location} occurrences at the
same path within 3 records~-- with e.g.~$\OptionalInventThres=50\%$
enough to infer the \type{location} attribute without original
annotation.
On the upper right of Figure \ref{fig:rightmove-example-aa}, we show
how the noisy price in the third record is eliminating: Again, the
characteristic tag paths are shown, leading to a support of $1/4$~--
with e.g.~$\OptionalDeleteThres=30\%$ too low to keep the
\type{bedRoomNumber} attribute.
The resulting data area and record layout is shown in the bottom of
Figure~\ref{fig:rightmove-example-aa}.

\section{Building the Domain Knowledge}
\label{sec:learning}

In \AMBER we assume that the domain schema is provided upfront by the
developer of the wrapper.
%
%
In particular, for a given extraction task, the developer must specify
only the schema $\SCHEMA=\REGATTS\cup\OPTATTS$ of regular and optional
attribute types, using the regular attribute types as strong
indicators for the presence of a \SCHEMA entity on a webpage.
In addition, the developer can also specify disjointness constraints
$\rho_1 \wedge \rho_2 \rightarrow \bot$ for two attribute types
$\rho_1, \rho_2\in\SCHEMA$ to force the domains of $\rho_1$ and
$\rho_2$ to be disjoint.

As mentioned earlier, devising basic gazetteers and regular
expressions
for core entities of a given domain requires very little work thanks
to frameworks like GATE~\cite{Cun11} and openly available knowledge
repositories such as DBPedia~\cite{Auer07dbpedia:a} and
FreeBase~\cite{Bollacker08}.
Values that can be recognised with regular expressions are usually
known a priori, as they correspond to common-sense entities, e.g.,
phone numbers or monetary values.
%
%
On the other hand, the construction of gazetteers, i.e., sets of terms
corresponding to the domains for attribute types (see
Section~\ref{sec:approach-def}), is generally a tedious task.
While it is easy to construct an initial set of terms for an attribute
type, building a complete gazetteer often requires an exhaustive
analysis of a large sample of relevant web pages.
Moreover, the domains of some attribute types are constantly changing,
for example a gazetteer for song titles is outdated quite quickly. 
Hence, in the following, we focus on the automatic construction and
maintenance of gazetteers and show how \AMBER's repeated-structure
analysis can be employed for growing small initial term sets into
complete gazetteers.

This automation lowers the need and cost for domain experts in the
construction of the necessary domain knowledge, since even a
non-expert can produce basic gazetteers for a domain to be completed
by our automated learning processes.
Moreover, the efficient construction of exhaustive gazetteers is
valuable for other applications outside web data extraction, e.g., to
improve existing annotation tools or to publish them as linked open
data for public use.

But even if a gazetteer is curated by a human, the resulting
annotations might still be noisy due to errors or intrinsic ambiguity
in the meaning of the terms.
Noise-tolerance is therefore of paramount importance in repairing or
discarding wrong examples, given enough evidence to support the
correction.
To this end, \AMBER uses the repeated structure analysis to infer
missing annotations and to discard noisy ones, incrementally growing
small seed lists of terms into complete gazetteers, and proving that
sound and complete initial domain knowledge is, in the end,
unnecessary.

\begin{algorithm}[tbp]\small
\SetKw{add}{add}
\SetKwFunction{components}{components}
\SetKwFunction{evidence}{ev}
\SetKwFunction{freq}{freq}
\SetKwData{Terms}{Terms}

\Input{$\TYPING$ -- extraction typing for $P$}
\Input{$\ANNOT$ -- annotations for $P$}
\Input{$\mathcal{U}=\{\UNIVERSE_1, \ldots, \UNIVERSE_k\}$ -- domains of the attribute types}
\Modifies{$\mathcal{U}$ -- modifies domains of attribute types}
\ForEach{$\left<n,\rho\right> \in \TYPING$}{\label{alg:learn:termextraction}
\Terms$\gets$ \components($n$)\;
\ForEach{$v \in \Terms\ |\ v \not \in \overline{\UNIVERSE}_{\rho}$}{
\If{$\left<\rho, n, v \right> \not \in \ANNOT$}{\label{alg:learn:termlearning}
\add $v$ to $\UNIVERSE_{\rho}$\;
$\evidence(v,\UNIVERSE_{\rho}) \gets \evidence(v,\overline{\UNIVERSE}_{\rho}) \gets 0$\;
}
$\evidence(v,\UNIVERSE_{\rho}) \gets \evidence(v,\UNIVERSE_{\rho}) + \freq^{+}(v,\rho,\TYPING)$\;\label{alg:learn:termreinforce}
}
}
\ForEach{$\ANNOT(\rho, n, v) | \left<n,\rho\right> \not \in \TYPING$}{
\label{alg:learn:termpruning}
$\evidence(v,\overline{\UNIVERSE}_{\rho}) \gets \evidence(v,\overline{\UNIVERSE}_{\rho}) + \freq^{-}(v,\rho,\TYPING)$\;
\If{$\evidence(v,\UNIVERSE_{\rho}) < \Theta \cdot \evidence(v, \overline{\UNIVERSE}_{\rho})$}{
\add $v$ to $\overline{\UNIVERSE}_{\rho}$\;
}
}
\caption{\learningalgo$(\TYPING, \ANNOT, \mathcal{U}$)}
\label{alg:learning}
\end{algorithm}

Learning in \AMBER can be carried in two different modes:
\begin{inparaenum}[\bfseries(1)]
\item In \emph{upfront learning}, \AMBER produces upfront domain
  knowledge for a domain to bootstrap the self-supervised wrapper
  generation.
\item In \emph{continuous learning}, \AMBER refines the domain
  knowledge over time, as \AMBER extracts more pages from websites of
  a given domain of previously unknown terms from nodes selected
  within the inferred repeated structure.
\end{inparaenum}
Regardless the learning mode, the core principle behind \AMBER's
learning capabilities is the mutual reinforcement of
repeated-structure analysis and the automatic annotation of the DOM of
a page.

For the sake of explanation, a single step of the learning process is
described in Algorithm~\ref{alg:learning}.
To update the gazetteers in $\mathcal{U}=\{\UNIVERSE_1, \ldots,
\UNIVERSE_k\}$ from an extraction typing \TYPING and the corresponding
annotations \ANNOT, for each node $n$, we compare the attribute types
of $n$ in \TYPING with the annotations \ANNOT for $n$.
This comparison leads to three cases:
\begin{asparaenum}[\bfseries(1)]
\item \emph{Term validation:} $n$ is a node attribute for $\rho$ and
  carries an annotation $\ANNOT(\rho,n,v)$.
  Therefore, $v$ was part of the gazetteer for $\rho$ and the
  repeated-structure analysis confirmed that $v$ is in the domain
  $\UNIVERSE_{\rho}$ of the attribute type.
\item \emph{Term extraction:} $n$ is a node attribute for $\rho$ but
  it does not carry an annotation $\ANNOT(\rho,n,v)$.
  Therefore, \AMBER should consider the terms in the textual content
  of $n$ for adding to the domain $\UNIVERSE_{\rho}$.
\item \emph{Term cleaning:} The node carries an annotation
  $\ANNOT(\rho,n,v)$ but does not correspond to an attribute node for
  $\rho$ in $\TYPING$, i.e., is \emph{noise} for $\rho$.
  Therefore, \AMBER must consider whether there is enough evidence to
  keep $v$ in $\UNIVERSE_{\rho}$.
\end{asparaenum}

For each attribute node $n$ in the extraction typing \TYPING, \AMBER
applies the function $\textsf{components}$ to tokenize the textual
content of the attribute node $n$ to remove unwanted token types
(e.g., punctuation, separator characters, etc.) and to produce a clean
set of tokens that are likely to represent terms from the domain.
For example, assume that the textual content of a node $n$ is the
string $w$=``Oxford, Walton Street, ground-floor apartment''.
The application of the function $\textsf{components}$ produces the set
$\langle$ ``Oxford'', ``Walton Street'', ``ground-floor'',
``apartment''$\rangle$ by removing the commas from $w$.

\AMBER then iterates over all terms that are not already known to
occur in the complement $\overline{\UNIVERSE}_{\rho}$ of the domain of
the attribute type $\rho$ and decides whether it is necessary to
validate or add them to the set of known values for $\rho$.
A term $v$ is in $\overline{\UNIVERSE}_{\rho}$ if is either known from
the schema that $v \in \UNIVERSE_{\rho^{\prime}}$ and $\SCHEMA \models
\rho \wedge \rho^{\prime} \rightarrow \bot$, or $v$ has been
recurrently identified by the repeated-structure analysis as noise.
Each term $v$ has therefore an associated value
$\textsf{ev}(v,\UNIVERSE_{\rho})$
(resp. $\textsf{ev}(v,\overline{\UNIVERSE}_{\rho})$) representing the
evidence of $v$ appearing --- over multiple learning steps --- as a
value for $\rho$ (resp. as noise for $\rho$).

If \AMBER determined that a node $n$ is an attribute node of type
$\rho$ but no corresponding annotation $\ANNOT(\rho,n,v)$ exists, then
we add them to the domain $\UNIVERSE_{\rho}$.
Moreover, once the term $v$ is known to belong to $U_{\rho}$ we simply
increase its evidence by a factor $\textsf{freq}^{+}(v, \rho,
\TYPING)$ that represent how frequently $v$ appeared as a value of
$\rho$ in the current extraction typing $\TYPING$.
The algorithm then proceeds to the reduction of the noise in the
gazetteer by checking those cases where an annotation
$\ANNOT(\rho,n,v)$ is not associated to any attribute node in the
extraction typing, i.e., it is noise for $\rho$.
Every time a term $v$ is identified as noise we increase the value of
$\textsf{ev}(v,\overline{\UNIVERSE}_{\rho})$ of a factor
$\textsf{freq}^{-}(v, \rho, \TYPING)$ that represents how frequently
the term $v$ occur as noise in the current typing $\TYPING$.
To avoid the accumulation of noise, \AMBER will permanently add a term
$v$ to $\overline{\UNIVERSE}_{\rho}$ if the evidence that $v$ is noisy
for $\rho$ is at least $\Theta$ times larger that the evidence that
$v$ is a genuine value for $\rho$.
The constant $\Theta$ is currently set to 1.5.

\begin{figure*}[tbp]
  \centering 
  \includegraphics[width=0.95\linewidth]{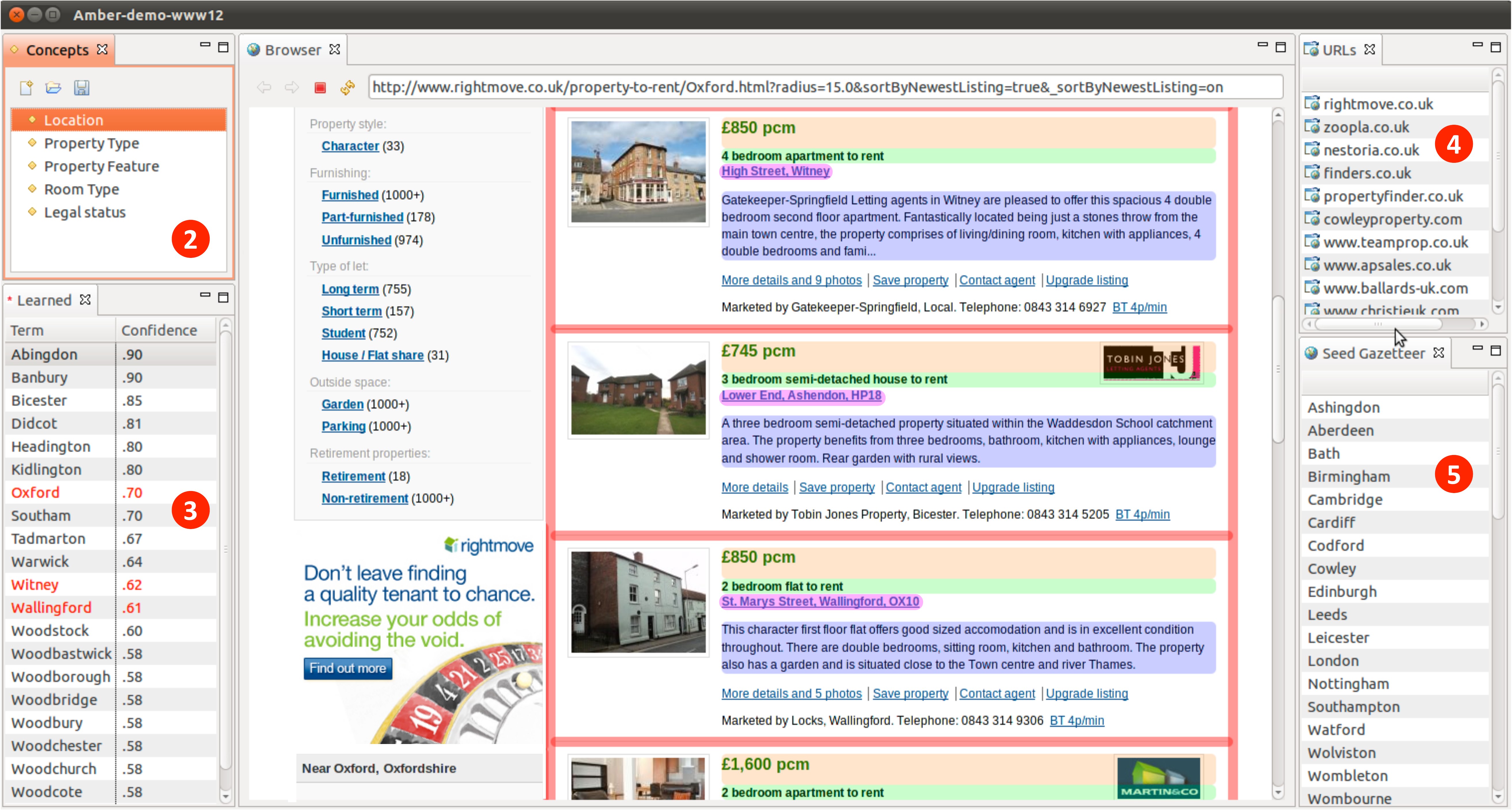} 
  \caption{\AMBER Learning Interface}
  \label{fig:amber-learning-gui}
\end{figure*}

To make the construction of the gazetteers even smoother, \AMBER also
provides a graphical facility (see
Figure~\ref{fig:amber-learning-gui}) that enables developers to
understand and possibly drive the learning process.
\AMBER's visual component provides a live graphical representation of
the result of the repeated-structure analysis on individual pages and
the position of the attributes \textcolor{red}{(1)}.  \AMBER relates
the concepts of the domain schema \textcolor{red}{(3)}, e.g.,
\TYPE{location} and \TYPE{property-type}, with \textcolor{red}{(3)}
the discovered terms, providing also the corresponding confidence
value.
The learning process is based on the analysis of a selected number of pages from  a list of URLs \textcolor{red}{(4)}.
The terms that have
been identified on the current page and have been validated are added to the gazetteer \textcolor{red}{(5)}.

\section{System Architecture}
\label{sec:system}
\begin{figure}[tbp]
  \centering
  \includegraphics[width=1\columnwidth]{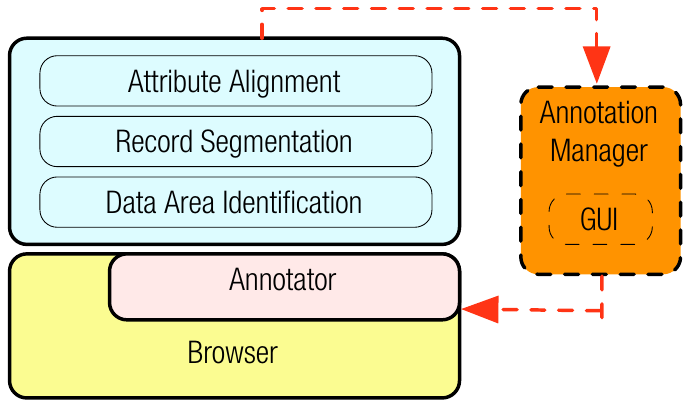}
  \caption{System architecture}
  \label{fig:architecture}
\end{figure}

Figure~\ref{fig:architecture} shows \AMBER's architecture composed of mainly of three layers.
The \textit{Browser Layer} consists of a JAVA API that abstracts the specific browser implementation actually employed. 
Through this API, currently \AMBER supports a real browser like Mozilla Firefox, as well as a headless browser emulator like HTMLUnit.
\AMBER uses the browser to retrieve the web page to analyze, thus having direct access to its DOM structure.
Such DOM tree is handed over to the \textit{Annotator Layer}. This is implemented such that different annotators can be plugged in and used
in combination, regardless their actual nature, e.g., web-service or custom standalone application. 
Given an annotation schema for the domain at hand, such layer produces annotations on the input DOM tree using all registered annotators.
Further, the produced annotations are reconciliated w.r.t. constraints present in the annotation schema.
Currently, annotations in \AMBER are performed by using a simple GATE (\url{gate.ac.uk}) pipeline consisting of gazetteers of terms and transducers (JAPE rules).
Gazetters for real estate and used cars domains are either manually-collect (for the most part) or derived from external sources
such as DBPedia and Freebase. Note that many types are common across domains (e.g., price, location, date), 
and that the annotator layer allows for arbitrary entity recognisers or annotators to be integrated.

With the annotated DOM at hand, \AMBER can begin its analysis with data area identification, record segmentation and attribute alignments.
Each of these phases is a distinct sub-module, and all of them are implemented in Datalog rules 
on top of a logical representation of the DOM and its annotations. These rules are with finite domains 
and non-recursive aggregation, and executed by the engine DLV. 

As described in Section~\ref{sec:approach}, the outcome of this analyses is an extraction typing $\TYPING$ along with 
attributes and relative support. During \AMBER's bootstrapping, however, $\TYPING$  is in turn used as feedback to realize the
learning phase (see Sect.~\ref{sec:learning}, managed by the \textit{Annotation Manager} module. 
Here, positive and negative lists of candidate terms is kept per each type, and used to update the initial gazetteers lists.
The Annotation Manager is optionally complemented with a graphical user interface, implemented as an Eclipse plugin (\url{eclipse.org})
which embeds the browser for visualization.

\newcommand{\kw}[1]{{\ensuremath {\mathsf{#1}}}\xspace}
\newcommand{\Prec}{\kw{Precision}}
\newcommand{\Recall}{\kw{Recall}}
\newcommand{\FScore}{\kw{FScore}}

\section{Evaluation}
\label{sec:evaluation}


\AMBER is implemented as a three-layer analysis engine where
\begin{inparaenum}[\bfseries(1)]
\item the \emph{web access layer} embeds a real browser to access and interact with the \emph{live} DOM
  of web pages,
\item the \emph{annotation layer} uses GATE~\cite{Cun11} along with domain gazetteers to
  produce annotations, and
\item the \emph{reasoning layer} implements the actual \AMBER algorithm as
  outlined in Section~\ref{sec:approach} in datalog rules over finite domains with
  non-recursive aggregation.
\end{inparaenum}

\subsection{\AMBERnofont in the UK}
\label{sec:qual-eval}

We evaluate \AMBER on $150$ UK real-estate web sites, randomly
selected among $2\,810$ web sites named in the yellow pages, and $100$
UK used car dealer websites, randomly selected from UK's largest used
car aggregator \url{autotrader.co.uk}. To assure diversity in our corpus, in case two
sites use the same template, we delete one of them and randomly choose another one.
For each site, we obtain one, or if possible, two result pages with at
least two result records.
These pages form the gold standard corpus, that is manually annotated
for comparison with \AMBER. For the UK real estate, the corpus
contains $281$ pages with $2\,785$ records and $14\,614$
attributes. The used car corpus contains $151$ pages with $1\,608$
records and $12\,732$ attributes.



\begin{figure}[tbp]
  \centering           
  \subfloat[Real Estate]{\label{fig:re-overall}\includegraphics[width=.25\textwidth]{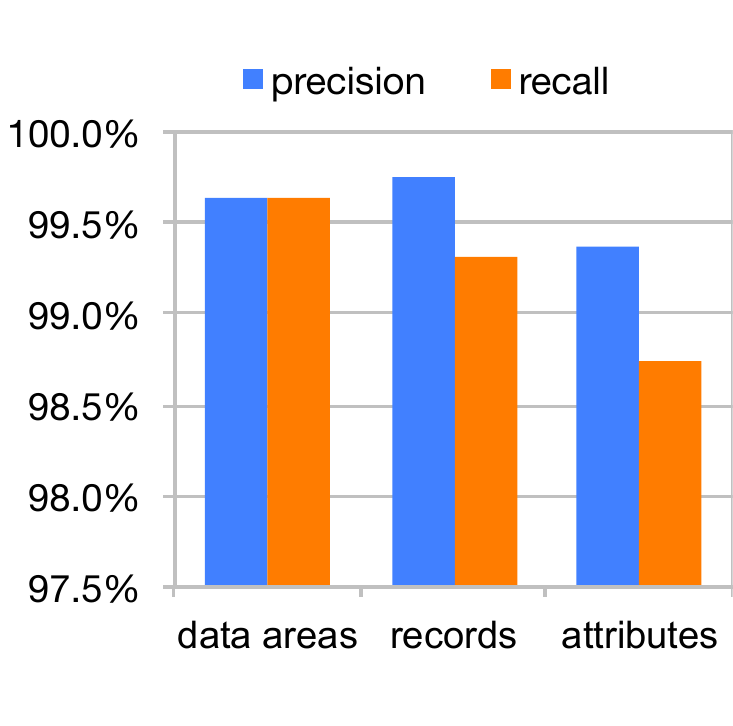}}
  \subfloat[Used Cars]{\label{fig:uc-overall}\includegraphics[width=.25\textwidth]{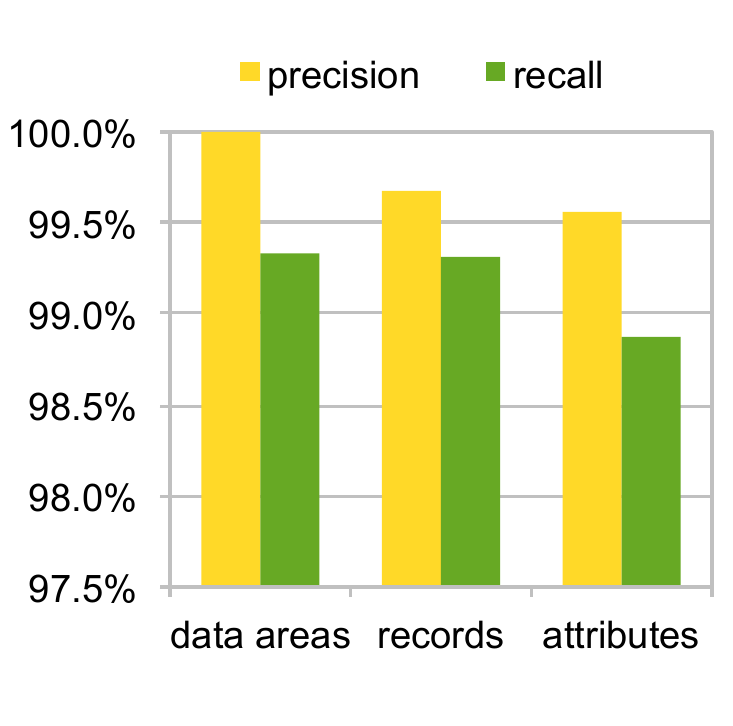}}
  \caption{Evaluation Overview}
  \label{fig:re-uc-overview}
\end{figure}

%

\begin{figure}[tbp]
  \centering
  \includegraphics[width=\columnwidth]{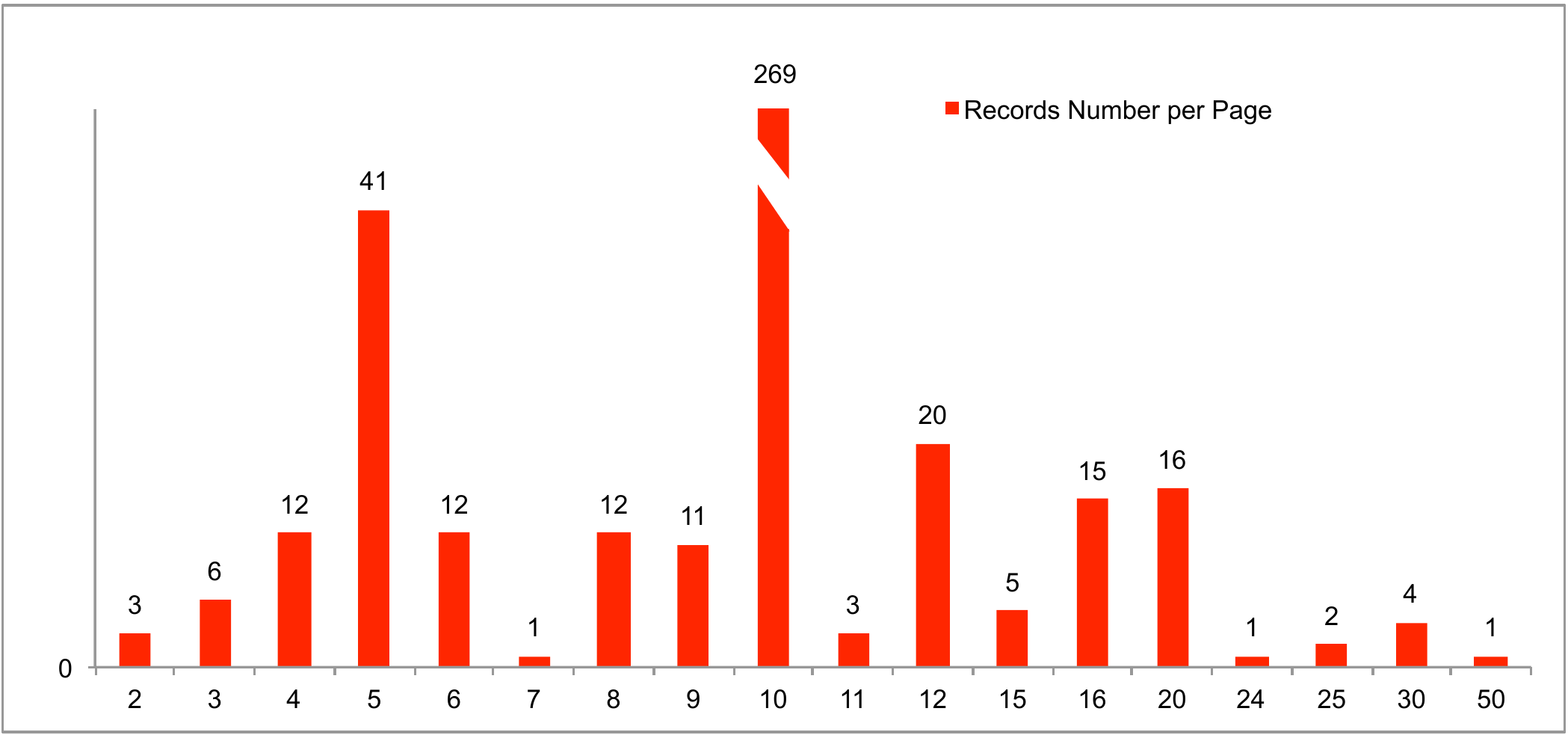}
  \caption{Distribution of Records Per Page}
  \label{fig:records-dist}
\end{figure}

\begin{figure}[tbp]
  \centering
  \includegraphics[width=\columnwidth]{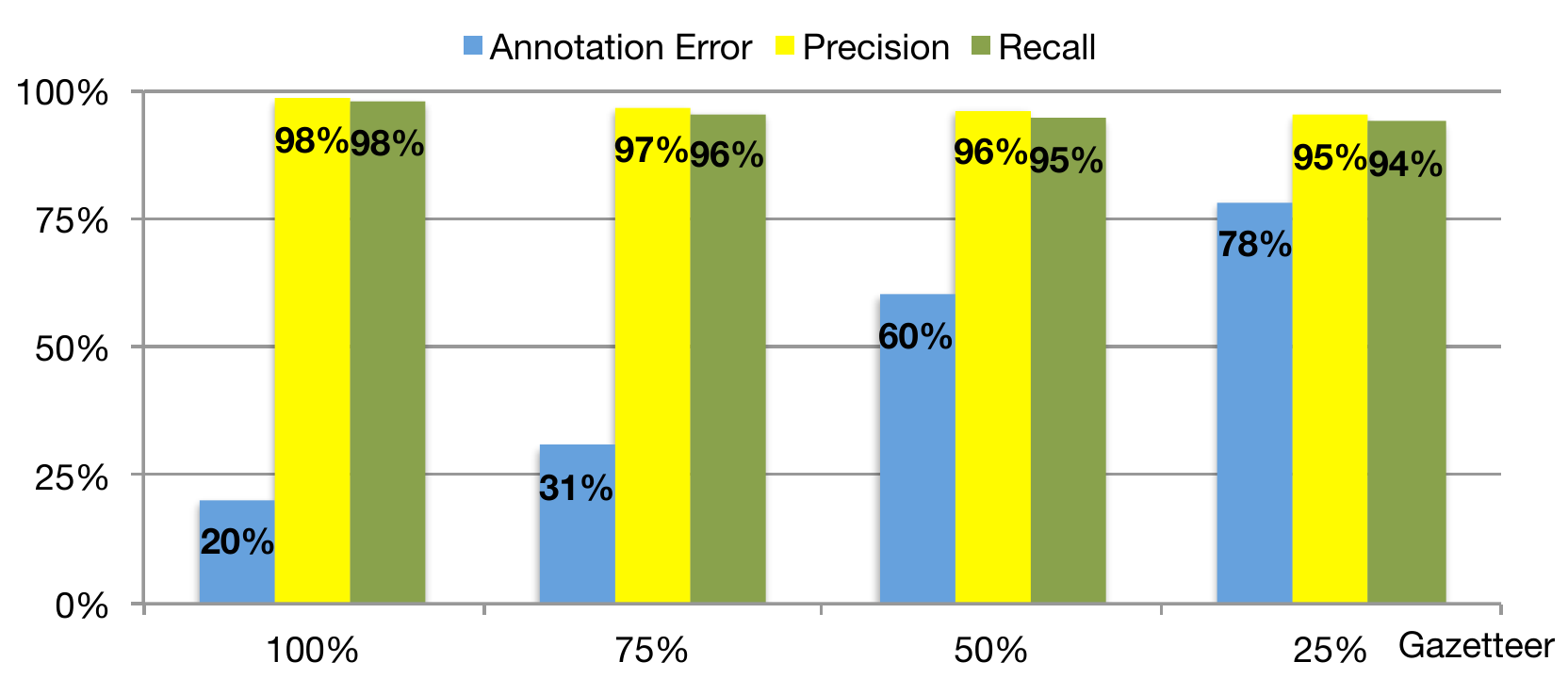}
  \caption{\AMBER Robustness wrt.\@ Noise in Gazetteers}
  \label{fig:robustness}
\end{figure}

%
%
%

For the following evaluations we use threshold values as
$\DEPTHTHRES=1$, $\DISTTHRES=2$ and $\RegularInventThres = 
\OptionalInventThres = 50\%$, $\RegularDeleteThres = 0\%$, and $\OptionalDeleteThres
= 20\%$. 
Figures~\ref{fig:re-overall} and~\ref{fig:uc-overall} show the overall precision
and recall of \AMBER on the real estate and used car corpora. As usual, precision is defined as the fraction of recognized data areas, records, or attributes that are
also present in the gold standard, whereas recall as the fraction of all data areas, records, and
attributes in the gold standard that is returned by \AMBER.
\AMBER achieves outstanding precision and recall on both domains ($>98\%$).
If we measure the average precision and recall per site (rather than the total
precision and recall), pages with fewer records have a higher impact. But even
in that harder case, precision and recall remains above $97.5\%$.  

\paragraph*{Robustness.}\ 
More importantly, \AMBER is very robust both w.r.t.\@ noise in the annotations/structure
and w.r.t.\@ the number of repeated records per page. 
To give an idea, in our corpus $50\%$ of pages contain structural noise either in the beginning or in the final part
of the data area. Also, $70\%$ of the pages contain noisy annotations for the  \TYPE{price} attribute, 
that is used as regular attribute in our evaluation.
On average, we count about 22 false occurrences per page. Nonetheless, \AMBER is able to perform nearly perfect accuracy,
fixing noise both from structure and annotations.  
Even worse, $100\%$ of pages contain noise for the \TYPE{Location} (i.e., addresses/locality, no postcode) attribute, 
which on average amounts to more than 50 (false positive) annotations of this type per page. 
To demonstrate how \AMBER copes with noisy annotations, we show in Figure~\ref{fig:robustness} the correlation between the noise levels (i.e., errors and incompleteness in the annotations) and \AMBER's performance in the extraction of the \TYPE{location} attribute. Even by using the full list of locations, about $20\%$ of all annotations are missed by the annotators, yet \AMBER achieves $>98\%$ precision and
recall. If we restrict the list to $75\%, 50\%$, and finally just $25\%$ of the
original list, the error rate rises over $30\%$ and $60\%$ to
$78\%$. Nevertheless, \AMBER's accuracy remains nearly unaffected dropping by
only $3\%$ to about $95\%$ (measuring here, of course, only the accuracy of
extraction location attributes). In other words, despite only getting
annotations for one out of every five locations, \AMBER in able to infer the other
locations from the regular structure of the records. \AMBER remains robust even if we introduce
errors for more than one attribute, as long as there is one regular attribute
such as the price for which the annotation quality is reasonable.  This
distinguishes \AMBER from all other approaches based on automatic annotations
that require reasonable quality (or at least, reasonable recall). \AMBER,
achieves high performance even from very poor quality annotators that can be
created with low effort.

At the same time, \AMBER is very robust w.r.t.\@ the number of records per
page. Figure~\ref{fig:records-dist} illustrates the distribution of record numbers per
page in our corpora. They mainly range from 4 to 20 records per page, with peaks
for 5 and 10 records. \AMBER performs
well on both small and large pages. Indeed, even in the case of only 3 records, 
it is able to exploit the repeated structure to achieve the correct extraction.

\begin{figure}[tbp]
  \centering
 \includegraphics[width=.85\columnwidth]{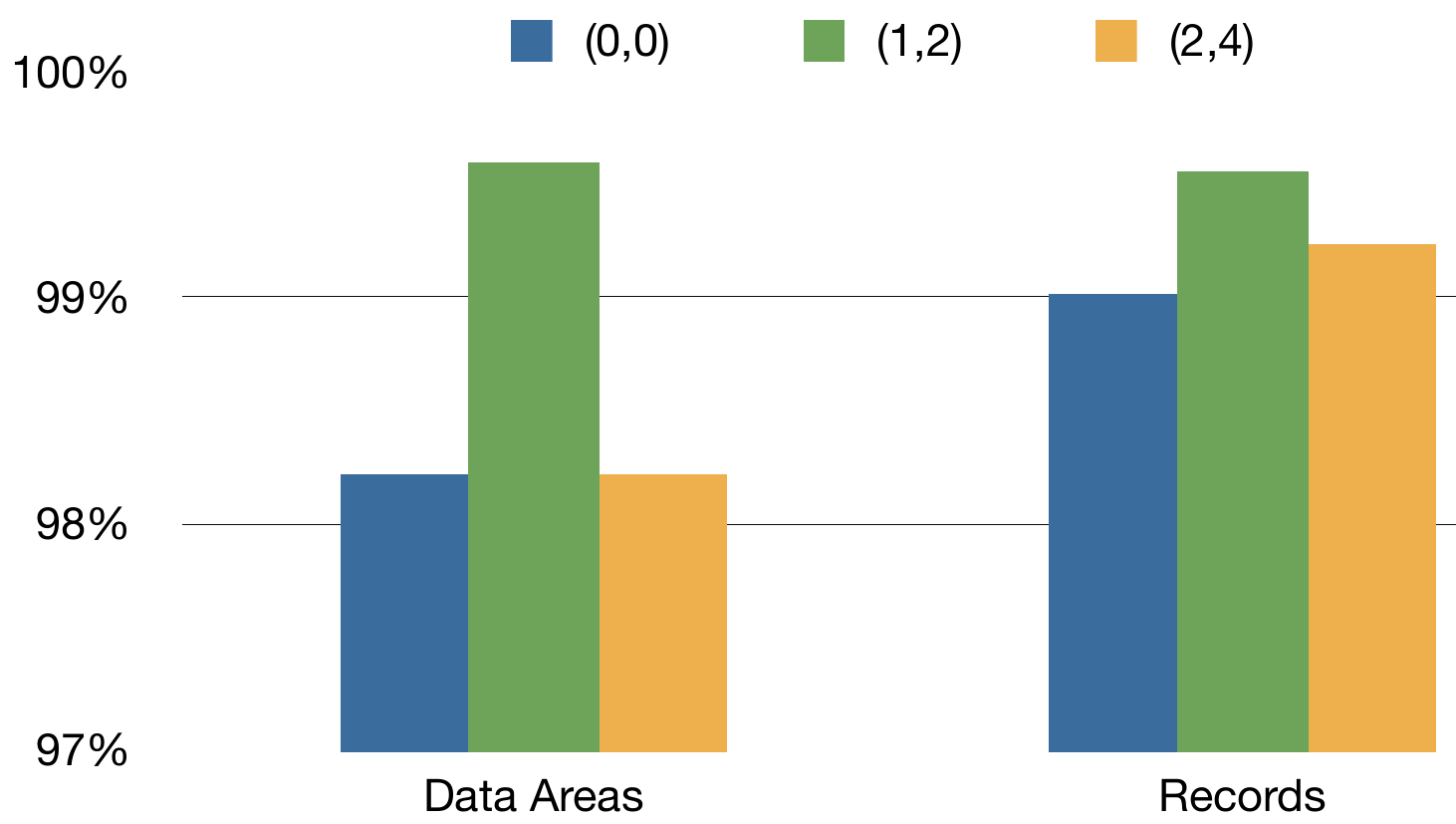}
  \caption{Depth/Distance Thresholds ($\DEPTHTHRES$,$\DISTTHRES$)}
  \label{fig:thresholds}
\end{figure}

Distance, depth, and attribute alignment thresholds can influence the performance of \AMBER. 
However, it is straightforward to choose good default values for these. For instance, considering 
the depth and distance thresholds, Figure~\ref{fig:thresholds} shows that the pair
($\DEPTHTHRES=1$,$\DISTTHRES=2$) provides significantly better
performance than ($0$,$0$) or ($2$,$4$).

\begin{figure}[tbp]
  \centering           
   \hspace*{-1em}
  \subfloat[Accuracy per Type]{\label{fig:attr_err}\includegraphics[width=.25\textwidth]{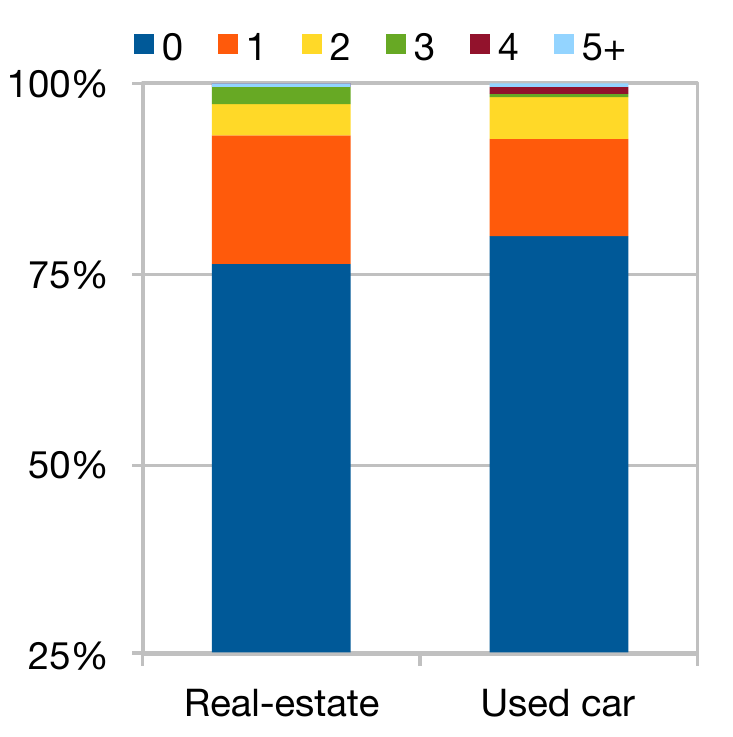}} 
  \subfloat[Accuracy per Number]{\label{fig:attr_errors100}\includegraphics[width=.25\textwidth]{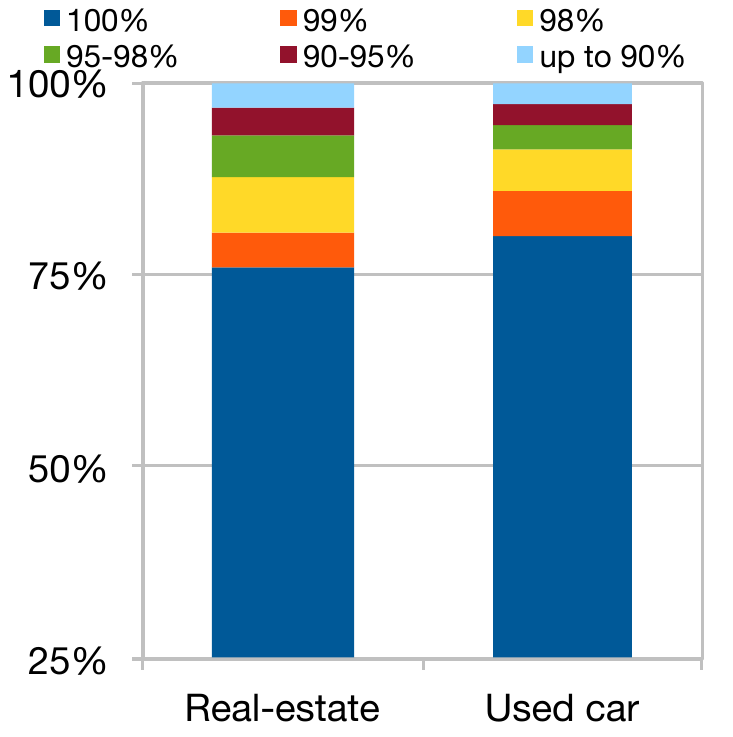}}  
   \caption{Attribute Extraction}
 \label{fig:attr-errors}
\end{figure}

\paragraph*{Attributes.}\ 
As far as attributes are concerned, there are $9$ different types for the real
estate domain, and $12$ different types for the used car corpus.
First of all, in $96\%$ of cases \AMBER perfectly recognizes objects, 
i.e., properly assigns all the attributes to the belonging object. 
It mistakes one attribute in $2\%$ of cases, and 2 and 3 attributes only in $1\%$ of cases, respectively.

\begin{figure}[tbp]
 \centering           
   \hspace*{-1em}
  \includegraphics[width=\columnwidth]{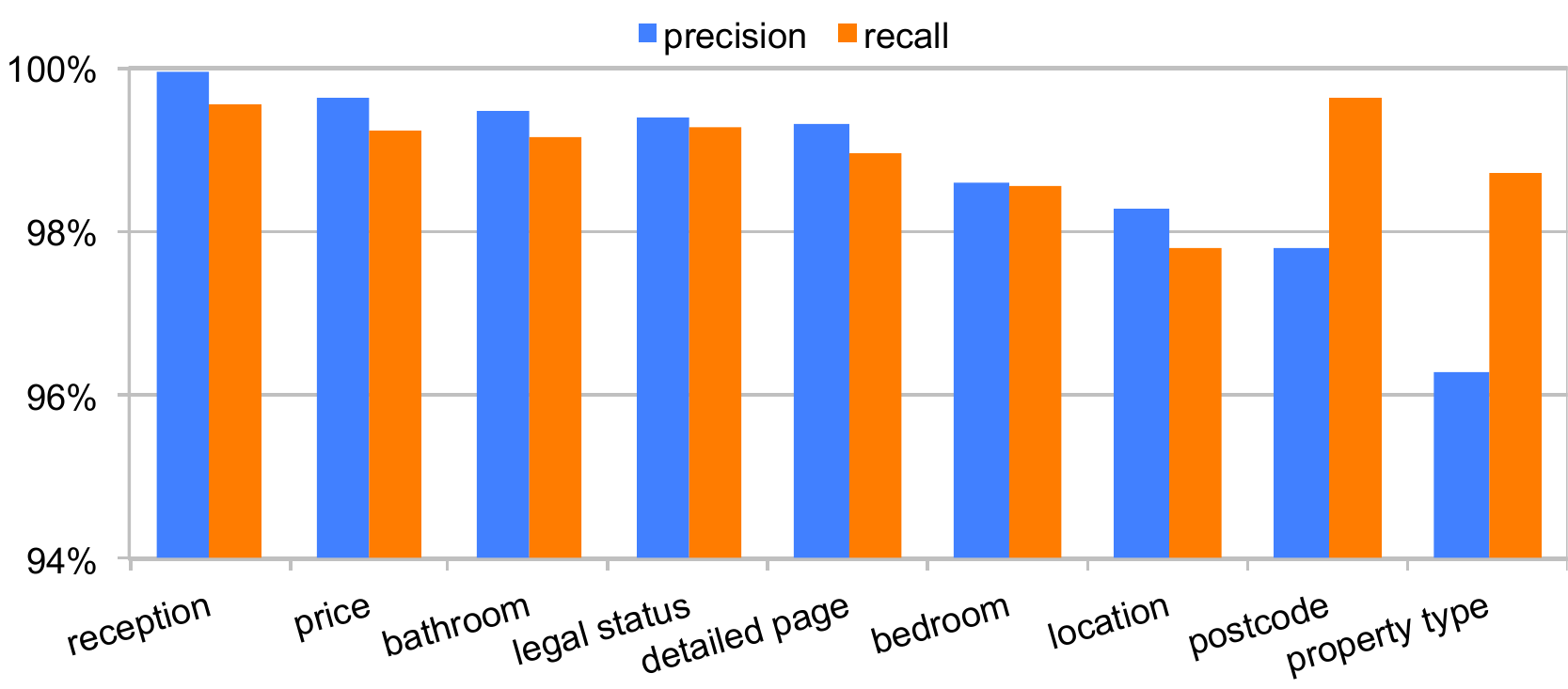}
   \caption{Real Estate Attributes Evaluation}
   \label{fig:re-attr-all}
\end{figure}

\begin{figure}[tbp]
  \centering
  \includegraphics[width=\columnwidth]{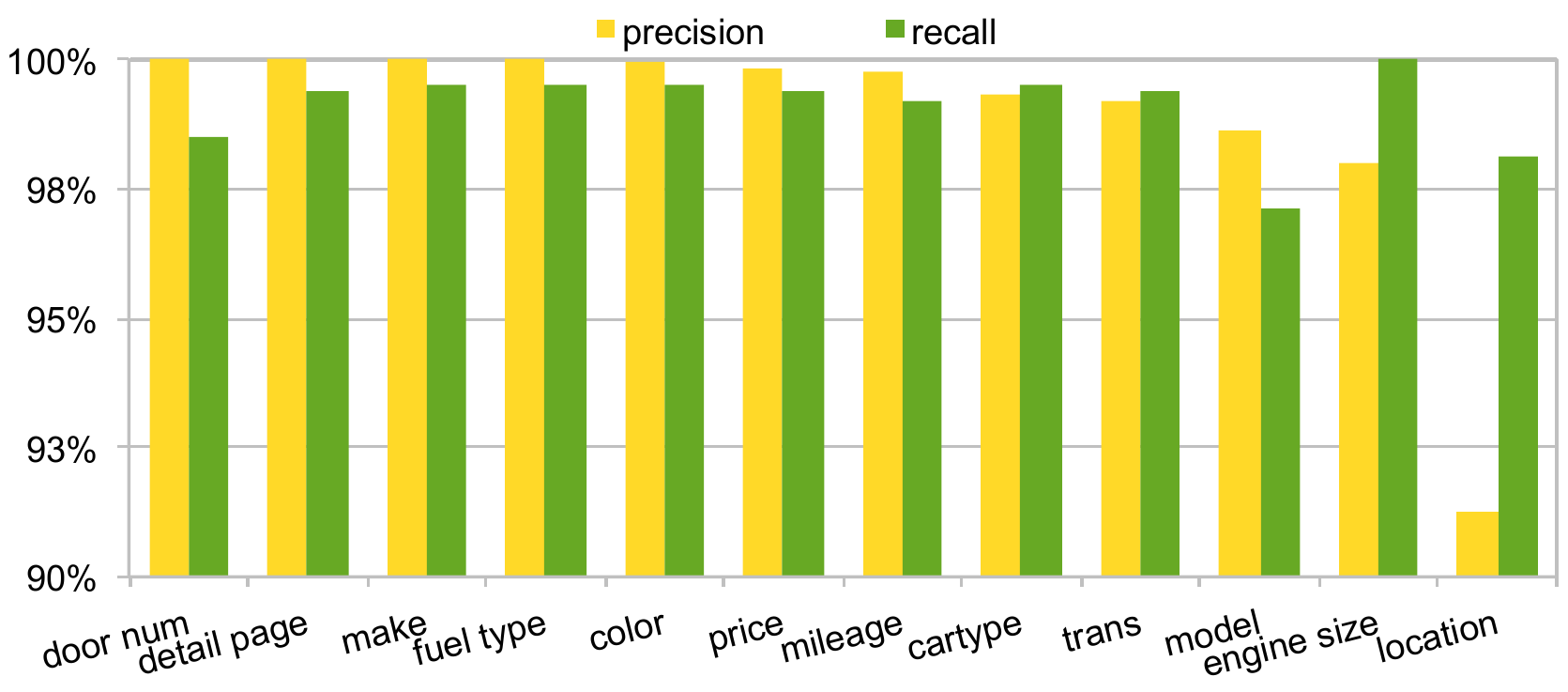}   
  \caption{Used Car Attributes Evaluation}
  \label{fig:uc-attr-all}
\end{figure}

Figure~\ref{fig:re-attr-all} illustrates the precision and recall that \AMBER achieves on each
individual attribute type of the real estate domain, where \AMBER reports nearly \emph{perfect recall} and very high precision ($>96\%$).
The results in the used car domain are similar (Figure~\ref{fig:uc-attr-all}) except for \TYPE{location}, where
\AMBER scores $91.3\%$ precision.  The reason is that, in this
particular domain, car models have a large variety of acronyms which happen to coincide with British postcodes, 
$e.g.$, N5 is the postcode of Highbury, London, X5 is a model of BMW, that also appear with regularity on the pages. 

Figures~\ref{fig:attr-errors} shows that on the vast majority of pages \AMBER achieves near
perfect accuracy.  Notably, in $97\%$ of cases, \AMBER retrieves correctly
between $90\%$ and $100\%$ of the attributes.
The percentage of cases in which \AMBER identifies
attributes from all attribute types is above 75\%, while only one type of attribute is wrong
in $17\%$ of the pages. For the remaining $6\%$ of pages \AMBER misidentifies
attributes from $2$ or $3$ types, with only one page in our corpora on which
\AMBER fails for $4$ attribute types. This emphasizes that on the vast majority
of pages at best one or two attribute types are problematic for \AMBER (usually
due to inconsistent representations or
optionality).

\subsection{Large-Scale Evaluation.}
\label{sec:quan-eval}
\begin{figure}[tbp]
  \centering
 \includegraphics[width=1\columnwidth]{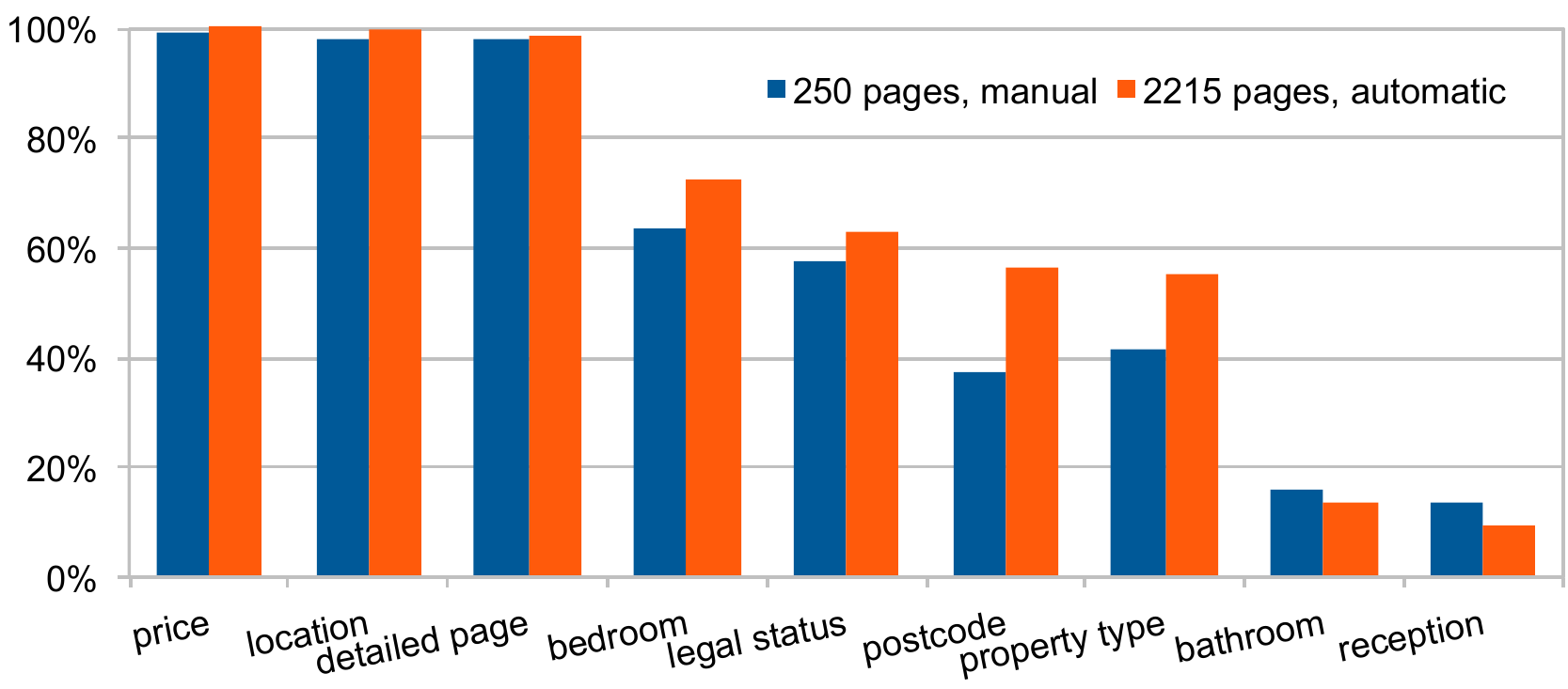}
  \caption{Attribute Frequencies in Large Scale Extraction}
  \label{fig:large-scale-fingerprint}
\end{figure}

To demonstrate \AMBER's ability to deal with a large set of diverse sites, we
perform an automated experiment beyond the sites catalogued in our gold
standard.
In addition to the $150$ sites in our real estate gold standard, we randomly
selected another $350$ sites from the $2\,810$ sites named in the yellow
pages. On each site, we manually perform a search until we reach the first
result page and retrieve all subsequent $n$ result pages and the expected number
of result records on the first $n-1$ pages, by manually counting the records on
the first page and assuming that the number of records remains constant on the
first $n-1$ pages (on the $n$th page the number might be smaller). This yields
$2\,215$ result pages overall with an expected number of $20\,723$ results
records. On this dataset, \AMBER identifies $20\,172$ records. 
Since a manual annotation is infeasible at this scale, we compare the frequencies of the individual types
of the extracted attributes with the frequencies of occurrences in the gold
standard, as shown in Figure~\ref{fig:large-scale-fingerprint}. Assuming that
both dataset are fairly representative selections of the whole set of result pages from
the UK real-estate domain, the frequencies of attributes should mostly coincide, as is the case in
Figure~\ref{fig:large-scale-fingerprint}. 
 Indeed, as shown in
 Figure~\ref{fig:large-scale-fingerprint} \TYPE{price}, \TYPE{location}, and
 \TYPE{details page} deviate by less than $2\%$, \TYPE{legal status},
 \TYPE{bathroom}, and \TYPE{reception number} by less than $5\%$. The high
 correlation strongly suggests that the attributes are mostly identified
 correctly. 
\TYPE{Postcode} and \TYPE{property type} cause a higher deviations of $18\%$ and
$12\%$, respectively. They are indeed attributes that are less reliably
identified by \AMBER, due to the reason explained above for UK postcodes and due
to the property type often appearing only within the free text property description. 

\subsection{Comparison with other Tools}
\label{sec:comp-with-other}

\paragraph*{Comparison with \ROADRUNNER.}\  
We evaluate \AMBER against \ROADRUNNER~\cite{crescenzi02:_roadr}, a fully
automatic system for web data extraction.  \ROADRUNNER does not extract
data areas and records explicitly, therefore we only compare the extracted
attributes.
\ROADRUNNER attempts to identify all repeated occurrences of variable data
(``slots'' of the underlying page template) and therefore extracts too many
attributes.
For example, \ROADRUNNER extracts on some pages more than $300$ attributes, mostly
URLs and elements in menu structures, where our gold standard contains only $90$
actual attributes. 
To avoid biasing the evaluation against \ROADRUNNER, we filter the output of
\ROADRUNNER, by
\begin{inparaenum}[]
\item removing the description block, 
\item duplicate URLs, and
\item attributes not contained in the gold standard, such as page or telephone
  numbers.
\end{inparaenum}


\begin{figure}[tbp]
  \centering
 \includegraphics[width=1\columnwidth]{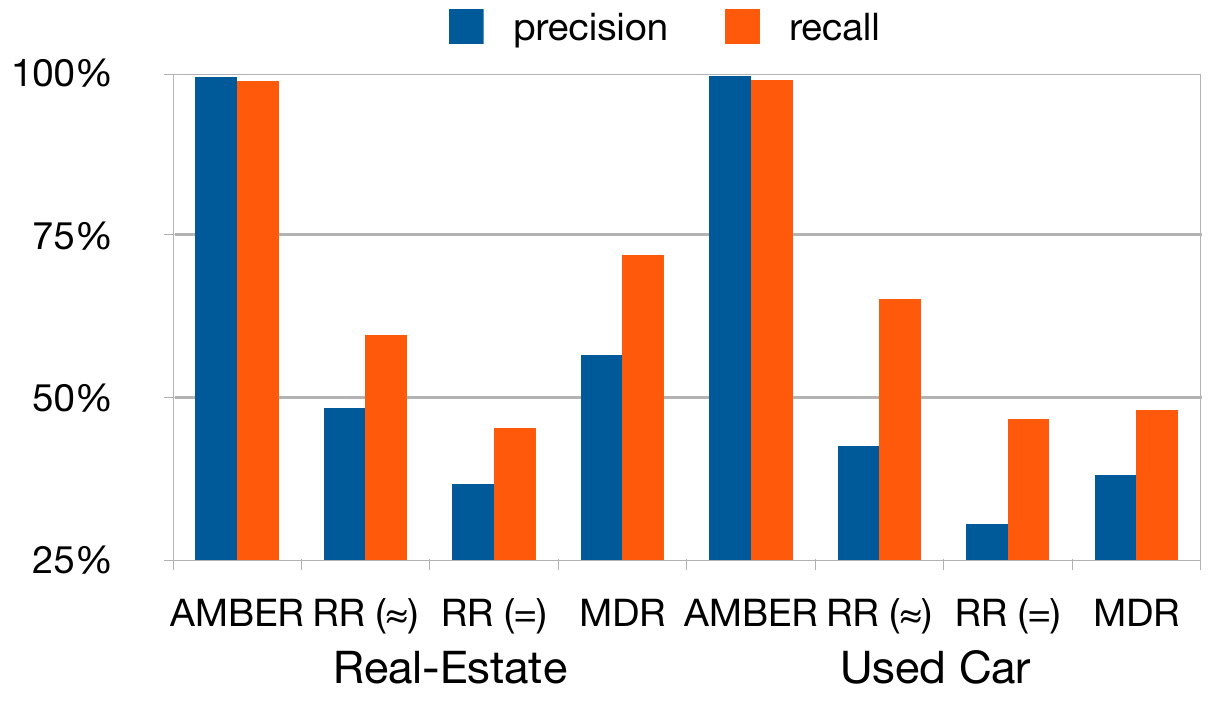}
  \caption{Comparison with \ROADRUNNER and \MDR}
  \label{fig:roadrunner-comparison}
\end{figure}


Another issue in comparing \AMBER with \ROADRUNNER is that \ROADRUNNER only
extracts entire text nodes. For example, \ROADRUNNER might extract ``Price
\pounds 114,995'', while \AMBER would produce ``\pounds 114,995''. Therefore we
evaluate \ROADRUNNER in two ways, once counting an attribute as correctly
extracted if the gold standard value is \emph{contained} in one of the
attributes extracted by \ROADRUNNER (RR $\approx$ in
Figure~\ref{fig:roadrunner-comparison}), and once counting an attribute only as
correctly extracted if the strings \emph{exactly} match (RR $=$ in
Figure~\ref{fig:roadrunner-comparison}).
Finally, as \ROADRUNNER works better with more than one result page from the
same site, we exclude sites with a single result page from this
comparison. The results are shown in
Figure~\ref{fig:roadrunner-comparison}. \AMBER outperforms \ROADRUNNER by
a wide margin, which reaches only $49\%$ in precision and $66\%$ in recall
compared to almost perfect scores for \AMBER. As expected, recall is
higher than precision in \ROADRUNNER.

\paragraph*{Comparison with \MDR.}\ 
We further evaluate \AMBER with \MDR, an automatic system for mining data records in web pages.
\MDR is able to recognize data areas and records, but unlike \AMBER, not attributes. 
Therefore in our comparison we only consider precision and recall for data areas and records
in both real estate and used cars domains. Also for the comparison with \ROADRUNNER, we
avoid biasing the evaluation against \MDR filtering out page portions e.g., menu, footer, pagination links, 
whose regularity in structure  misleads \MDR. Indeed, these are recognized by \MDR as data areas or records. 
Figure~\ref{fig:roadrunner-comparison} illustrates the results. In all cases, \AMBER outperforms \MDR which  
on used-cars reports $57\%$ in precision and $72\%$ in recall as best performance.
\MDR suffers the complex structure of data records, which may contain optional information as nested repeated structure.
This, in turn, are often (wrongly) recognized by \MDR as record (data area).

\subsection{\AMBERnofont Learning}
\label{sec:ambernofont-learning}

The evaluation of \AMBER's learning capabilities is done with respect to the upfront learning mode discussed in Section~\ref{sec:learning}. In particular, we want to evaluate \AMBER's ability of constructing an accurate and complete gazetteer for an attribute type from an incomplete and noisy \emph{seed} gazetteer. We show that at each learning iteration (see Algorithm~\ref{alg:learning} in Section~\ref{sec:learning}) the accuracy of the gazetteer is significantly improved, and that the learning process converges to a stable gazetteer after few iterations, even in the case of attribute types with large and/or irregular value distributions in their domains.

\paragraph*{Setting.}\ In the evaluation that follows we show \AMBER's learning behaviour on the \TYPE{location} attribute type. In our setting, the term location refers to formal geographical locations such as towns, counties and regions, e.g., ``Oxford'', ``Hampshire'', and ``Midlands''. Also, it is often the case that the value for an attribute type consists of multiple and somehow structured terms, e.g., ``The Old Barn, St. Thomas Street - Oxford''.
The choice of \TYPE{location} as target for the evaluation is justified by the fact that this attribute type has typically a very large domain consisting of ambiguous and severely irregular terms. Even in the case of UK locations alone, nearly all terms from the English vocabulary either directly correspond to a location name (e.g., ``Van'' is a location in Wales) or they are part of it (e.g., ``Barnwood'', in Gloucestershire).
The ground truth for the experiment consists of a clean gazetteer of 2010 UK locations and 1,560 different terms collected from a sample of 235 web pages sourced from 150 different UK real-estate websites.
\paragraph*{Execution.}\ We execute the experiment on two different seed gazetteers $G_{20}$ (resp. $G_{25}$) consisting of a random sample of 402 (resp. 502) UK locations corresponding to the $20\%$ (resp. $25\%$) of the ground truth.

\begin{table}[tbp]
  \centering
  \caption{Learning performance on $G_{20}$.}
  {\small
  \begin{tabular}{ccccccc}
    \toprule
    rnd. & L$^{E}$ & C$^{E}$ & P$^{E}$ & R$^{E}$ & L$^{L}$ & C$^{L}$ \\
    \midrule
    1 & 1009 & 763  & 75.65\% & 37.96\% & 169 & 147 \\ 
    2 & 1300 & 1063 & 81.77\% & 52.89\% & 222 & 196 \\
    3 & 1526 & 1396 & 91.48\% & 69.45\% & 224 & 205 \\
    4 & 1845 & 1773 & 96.10\% & 88.21\% & 59  & 52  \\
    5 & 1862 & 1794 & 96.35\% & 89.25\% & 23  & 19  \\
    6 & 1862 & 1794 & 96.35\% & 89.25\% & 0   & 0   \\
    \bottomrule
  \end{tabular}}  
  
  \label{tab:g-20}
\end{table}

By taking as input $G_{20}$, the learning process saturates (i.e., no new terms are learned or dropped) after six iterations with a $92.66\%$ accuracy (F$_1$-score), while with $G_{25}$, only 5 iterations are needed for an accuracy of $92.79\%$. Note that at the first iteration the accuracy is $50.54\%$ for $G_{20}$ and $60.94\%$ for $G_{25}$
Table~\ref{tab:g-20} and Table~\ref{tab:g-25} show the behaviour for each learning round. We report the number of locations extracted (L$^{E}$), i.e., the number of attribute nodes carrying an annotation of type \TYPE{location}; among these, C$^{E}$ locations have been correctly extracted, leading to a precision (resp. recall) of the extraction of P$^{E}$ (resp. R$^{E}$). The last two columns show the number of learned instances (L$^{L}$), i.e., those added to the gazetteer and, among these, the correct ones (C$^{L}$).

It is easy to see that the increase in accuracy is stable in all the learning rounds and that the process quickly converges to a stable gazetteer.

\begin{table}[tbp]
  \centering
  \caption{Learning performance on $G_{25}$.}
  {\small
  \begin{tabular}{ccccccccc}
     \toprule
    rnd. & L$^{E}$ & C$^{E}$ & P$^{E}$ & R$^{E}$ & L$^{L}$ & C$^{L}$ \\
    \midrule
    1 & 1216 & 983  & 80.84\% & 48.91\% & 289 & 248 \\ 
    2 & 1538 & 1334 & 86.74\% & 66.37\% & 225 & 204 \\
    3 & 1717 & 1617 & 94.18\% & 80.45\% & 57  & 55  \\
    4 & 1960 & 1842 & 93.98\% & 91.64\% & 44  & 35  \\
    5 & 1960 & 1842 & 93.98\% & 91.64\% & 0   & 0   \\
    \bottomrule
  \end{tabular}}
  \label{tab:g-25}
\end{table}

\section{Related Work}
\label{sec:related-work}

The key assumption in web data extraction is that a large fraction of
the data on the web is
structured~\cite{DBLP:journals/cacm/CafarellaHM11} by HTML markup and
visual styling, especially when web pages are automatically generated
and populated from templates and underlying information systems. This 
sets web data extraction apart from information extraction where entities,
relations, and other information are extracted from free text (possibly from web
pages).

Early web data extraction approaches address data extraction via manual wrapper
development \cite{hammer97:_semis_data} or through visual, semi-automated tools
\cite{Baumgartner2001VisualWebIEwithLixto,laender02:_debye} (still commonly used
in industry).  Modern web data extraction approaches, on the other hand,
overwhelmingly fall into one of two categories (for recent surveys, 
see~\cite{chang06:_survey_of_web_infor_extrac_system,565137}): \emph{Wrapper
  induction}~\cite{DBLP:conf/sigmod/DalviBS09,%
  freitag00:_machin_learn_for_infor_exrtr,%
  DBLP:conf/icde/GulhaneMMRRSSTT11,%
  hsu98:_gener_finit_state_trans_for,%
  JiLi10,%
  DBLP:journals/dke/KosalaBBB06,%
  kushmerick97:_wrapp_induc_for_infor_extrac,%
  muslea01:_hierar_wrapp_induc_for_semis_infor_system}
starts from a number of manually annotated examples, i.e., pages where the
objects and attributes to be extracted are marked by a human, and automatically
produce a wrapper program which extracts the corresponding content from
previously unseen pages.
\emph{Unsupervised wrapper generation}~\cite{crescenzi02:_roadr,%
  kayed10:_fivat,%
  liu06:_vision_based_web_data_recor_extrac,%
  simon05:_viper,%
  su09ode,%
  DBLP:conf/kdd/WangCWPBGZ09,%
  zhai06:_struc_data_extrac_from_web} 
attempts to fully automate the extraction process by unsupervised
learning of repeated structures on the page as they usually indicate
the presence of content to be extracted. 

Unfortunately, where the former are limited in automation, the latter are in
accuracy. This has caused a recent flurry of
approaches~\cite{DBLP:conf/vlds/CreoCQM12,%
  DBLP:journals/pvldb/DalviKS11,%
  DBLP:conf/icde/DerouicheCA12,%
  SMM*08} that like \AMBER attempt to automatise the production of 
examples for wrapper inducers through existing entity recognisers or similar
automatic annotators. Where these approaches differ most is how and to what
extend they address the inevitable noise in these automatic annotations.

%
%

\subsection{Wrapper Induction Approaches}
\label{sec:wrapp-induct-appr}

Wrapper induction 
can deliver highly accurate results provided correct and complete input
annotations. 
The process is based on the iterative generalization of properties
(e.g., structural and visual) of the marked content on the input
examples.
The learning algorithms infer generic and possibly robust extraction
rules in a suitable format, e.g., XPath
expressions~\cite{DBLP:conf/sigmod/DalviBS09,%
  DBLP:conf/icde/GulhaneMMRRSSTT11} or
automata~\cite{hsu98:_gener_finit_state_trans_for,%
  muslea01:_hierar_wrapp_induc_for_semis_infor_system}, that are
applicable to similar pages for extracting the data they are generated
from.

The structure of the required  example annotations differs across
different tools, impacting the complexity of the learned wrapper and
the accuracy this wrapper achieves.
%
%
Approaches such as~\cite{freitag00:_machin_learn_for_infor_exrtr,%
  DBLP:journals/dke/KosalaBBB06} operate on \emph{single attribute} annotations,
i.e., annotations on a single attribute or multiple, but a-priori unrelated
attributes.
As a result, the wrapper learns the extraction rules independently for
each attribute, but, in the case of multi-attribute objects, this
requires a subsequent reconciliation phase.
The approaches presented in~\cite{hsu98:_gener_finit_state_trans_for,%
  kushmerick97:_wrapp_induc_for_infor_extrac,%
  muslea01:_hierar_wrapp_induc_for_semis_infor_system} are based on
\emph{annotated trees}.
The advantage w.r.t. single-attribute annotations
is that tree annotations make easier to recognize nested structures.

By itself, wrapper induction is incapable of scaling to the web.
Because of the wide variation in the template structures of given web
sites, it is practically impossible to
annotate a sufficiently large page set
to cover all relevant combinations of features indicating the presence
of structured data.
%
%
More formally, the \emph{sample complexity} for web-scale supervised
wrapper induction is too high in all but some restricted cases, as
in e.g.~\cite{DBLP:conf/kdd/WangCWPBGZ09} which extracts news titles and
bodies.
Furthermore, traditional wrapper inducers are very sensitive to
incompleteness and noise in the annotations thus requiring considerable human
effort to create such low noise and complete annotations.


\subsection{Unsupervised Web Data Extraction}
\label{sec:unsup-web-data}

The completely unsupervised generation of wrappers has been based on discovering
regularities on pages presumably generated by a common template. 
Works such as ~\cite{kayed10:_fivat,%
  liu03:_minin_data_recor_in_web_pages,%
  liu06:_vision_based_web_data_recor_extrac,%
  simon05:_viper,%
  zhai06:_struc_data_extrac_from_web,%
  zhao05:_fully_autom_wrapp_gener_for_searc_engin,%
  zhao06:_autom_extrac_of_dynam_recor} discuss
\emph{domain-independent} approaches that only rely on repeated HTML
markup or regularities in the visual rendering.
The most common task that can be solved by these tools is \emph{record
  segmentation}~\cite{liu03:_minin_data_recor_in_web_pages,%
  zhao05:_fully_autom_wrapp_gener_for_searc_engin,%
  zhao06:_autom_extrac_of_dynam_recor}, where an area of the page is
segmented into regular blocks each representing an object to be
extracted.
Unfortunately, these systems are quite susceptible to noise in the repeated
structure as well as to regular, but irrelevant structures such as navigation
menus. This limits their accuracy severely, as also demonstrated in
Section~\ref{sec:evaluation}. 
%
In \AMBER, having domain specific annotators at hand, we also exploit
the underlying repeated structure of the pages, but guided by occurrences of
regular attributes which allow us to distinguish relevant data areas from noise,
as well as to address noise among the records. 
This allows us to extract records with higher precision.

A complementary line of work deals with specifically stylized
structures, such as
\emph{tables}~\cite{DBLP:journals/pvldb/CafarellaHWWZ08,Gatterbauer:2007:TDI:1242572.1242583}
and \emph{lists}~\cite{DBLP:journals/vldb/ElmeleegyMH11}.
The more clearly defined characteristics of these structures enable
domain-independent algorithms that achieve fairly high precision in distinguish
genuine structures with relevant data from structures created only for layout
purposes. They are particular attractive for use in settings such as web search
that optimise for coverage over all sites rather than recall from a particular
site.

Instead of limiting the structure types to be recognized, one can
exploit \emph{domain knowledge} to train more specific models.
\emph{Domain-dependent} approaches such as
\cite{DBLP:conf/kdd/WangCWPBGZ09,ZNW*06} exploit specific properties
for record detection and attribute labeling.
However, besides the difficulty of choosing the features to be
considered in the learning algorithm for each domain, changing the
domain usually results in at least a partial retraining of the models
if not an algorithmic redesign.

More recent approaches are, like \AMBER and the approaches discussed in
Section~\ref{sec:combined-approaches}, \emph{domain-parametric}, i.e., they
provide a domain-independent framework which is parameterized with a specific
application domain. 
For instance, \cite{su09ode} uses a domain ontology for data area
identification but ignores it during record segmentation.
%

\subsection{Combined Approaches}
\label{sec:combined-approaches}

Besides \AMBER, we are only aware of three other
approaches~\cite{DBLP:journals/pvldb/DalviKS11,DBLP:conf/icde/DerouicheCA12,SMM*08}
that exploit the mutual benefit of unsupervised extraction and induction from
automatic annotations. All these approaches are a form of \emph{self-supervised}
learning, a concept well known in the machine learning community and that has
already been successfully applied in the information extraction
setting~\cite{springerlink:10.1007/s10115-007-0110-6}.

In~\cite{SMM*08}, web pages are independently annotated using
background knowledge from the domain and analyzed for repeated
structures with conditional random fields (CRFs).
The analysis of repeated structures identifies the record structure in
searching for evenly distributed annotations to validate (and
eventually repair) the learned structure.
Conceptually, \cite{SMM*08} differs from \AMBER as it initially infers
a repeating page structure with the CRFs independently of the
annotations.
\AMBER, in contrast, analyses only those portions of the page that are
more likely to contain useful and regular data.
Focusing the analysis of repeated structures to smaller areas is
critical for learning an accurate wrapper since complex pages might
contain several regular structures that are not relevant for the
extraction task at hand.
This is also evident from the reported accuracy of the method proposed
in~\cite{SMM*08} that ranges between $63\%$ and $85\%$ on attributes,
which is significantly lower than \AMBER's accuracy.

This contrasts also with \cite{DBLP:journals/pvldb/DalviKS11} which aims at
making wrapper induction robust against noisy and incomplete annotations, such
that fully automatic and cheaply generated examples are sufficient.
The underlying idea is to induce multiple candidate wrappers by using
different subsets of the annotated input.
The candidate wrappers are then ranked according to a probabilistic
model, considering both features of the annotations and the page
structure.
This work has proven that, provided that the induction algorithm satisfies few
reasonable conditions, it is possible to produce very accurate wrappers for
single-attribute extraction, though sometimes at the price of hundreds of calls of
the wrapper inducer.
For multi-attribute extraction, \cite{DBLP:journals/pvldb/DalviKS11} reports
high, if considerably lower accuracy than in the single-attribute case. More
importantly, the wrapper space is considerably larger as the number of
attributes acts as a multiplicative factor. Unfortunately, no performance
numbers for the multi-attribute case are reported in
\cite{DBLP:journals/pvldb/DalviKS11}. 
In contrast, \AMBER fully addresses the problem of multi-attribute
object extraction from noisy annotations by
eliminating the annotation errors during the attribute alignment.
%
%
%
Moreover, \AMBER also avoids any assumptions on a subsequently
employed wrapper induction system.

A more closely-related work is
\tool{ObjectRunner}~\cite{DBLP:conf/icde/DerouicheCA12}, a tool driven by an
intensional description of the objects to be extracted (a SOD in the terminology
of~\cite{DBLP:conf/icde/DerouicheCA12}).
A SOD is basically a schema for a nested relation with attribute
types.
Each type comes with associated annotators (or recognizers) for
annotating the example pages to induce the actual wrapper from by a
variant of
\tool{ExAlg}~\cite{arasu03:_extrac_struc_data_from_web_pages}.
The SOD limits the wrapper space to be explored
($\leq$20 calls of the inducer) and improves the quality of the
extracted results. This is similar to \AMBER, though \AMBER not only limits the
search space, but also considers only alternative segmentations instead of full
wrappers (see Section~\ref{subsec:segmentation}). 
On the other hand, the SOD can seriously limit the recall of the
extraction process, in particular, since the matching conditions of a
SOD strongly privilege precision.
The approach is furthermore limited by the rigid coupling of attribute
types to separators (i.e., token sequences acting act as boundaries
between different attribute types).
It fact attribute types appear quite frequently together with very
diverse separators (e.g., caused by a special highlighting or by a
randomly injected advertisement).
The process adopted in \AMBER is not only tolerant to noise in the annotations
but also to random garbage content between attributes and between records as it
is evident from the results of our evaluation: Where \tool{ObjectRunner} reports
that between $65\%$ and $86\%$ of the objects in $5$ domains ($75\%$ in the car
domain) are extracted without any error, \AMBER is able to extract over
$95\%$ of the objects from the real estate and used car domain without any
error.

\section{Conclusion}
\label{sec:conclusion}

\AMBER pushes the state-of-the-art in extraction of multi-attribute objects from
the deep web, through a fully-automatic approach that combines the analysis of the repeated
structure of the web page and automatically-produced annotations. \AMBER compensates
for noise in both the annotations and the repeated structure to achieve $>98\%$
accuracy for multi-attribute object extraction. To do so, \AMBER requires a
small amount of domain knowledge that can is proven (Section~\ref{sec:learning}) to be easily obtainable from just a few example instances and pages. 

Though \AMBER is outperforming existing approaches by a notable margin for
multi-attribute extraction on product domains, there remain a number of open
issues in \AMBER and multi-attribute object extraction in general:
\begin{asparaenum}[\bfseries(1)]
\item \textbf{Towards irregular, multi-entity domains.}  Domains with
  \emph{multiple entity types} have not been a focus of data extraction systems
  in the past and pose a particular challenge to approaches such as \AMBER that
  are driven by domain knowledge. While dealing with the (frequent) case in
  which these heterogeneous objects share a common regular attribute is fairly
  straightforward, more effort it is necessary when regular attributes are
  diverse.  To this end, more \emph{sophisticated} regularity conditions may be
  necessary. Similarly, the ambiguity of instance annotators may be so
  significant that a stronger reliance on labels in structures such as tables is
  necessary. 

\item \textbf{Holistic Data Extraction.}  Though data extraction involves
  several tasks, historically they have always been approached in isolation.  
  Though some approaches have considered form understanding and extraction from
  result pages together, a truly holistic approach that tries to reconcile
  information from forms, result pages, details pages for individual objects,
  textual descriptions, and documents or charts about these objects remains
  an open challenge.


\item \textbf{Whole-Domain Database.}  \AMBER, as nearly all existing data
  extraction approaches, is focused on extracting objects from a given site.
  Though unsupervised approaches such as \AMBER can be applied to many sites,
  such a domain-wide extraction also requires data integration between sites and
  opens new opportunities for cross-validation between domains. In particular, 
  domain-wide extraction enables automated learning not only for instances as in
  \AMBER, but also for new attributes through collecting sufficiently large sets
  of labels and instances to use ontology learning approaches. 
\end{asparaenum}

\section{Acknowledgements}
\ACKNOWLEDGEMENTS

\def\Nst#1{$#1^{st}$}\def\Nnd#1{$#1^{nd}$}\def\Nrd#1{$#1^{rd}$}\def\Nth#1{$#1^{th}$}

\end{document}